# Accounting for skill in trend, variability, and autocorrelation facilitates better multi-model projections: application to the AMOC and temperature time series

Projections accounting for model trend and variability skill

Roman Olson[1,2,3], Soon-Il An[1,*], Yanan Fan[4] and Jason P. Evans[5]

[1]Department of Atmospheric Sciences, Yonsei University, Seoul, South Korea

[2]Center for Climate Physics, Institute for Basic Science, Busan, South Korea

[3]Pusan National University, Busan, South Korea

[4]School of Mathematics and Statistics, UNSW Australia, Sydney, NSW, Australia

[5]Climate Change Research Centre and ARC Centre for Excellence in Climate Extremes, UNSW Australia, Sydney, NSW, Australia

**Contents of supporting information S1 file follow the main text. See the published version to access the contents of the supporting information S2 file.**

[*]Corresponding author

E-mail: sian@yonsei.ac.kr (S-I A)

# Abstract

We present a novel quasi-Bayesian method to weight multiple dynamical models by their skill at capturing both potentially non-linear trends and first-order autocorrelated variability of the underlying process, and to make weighted probabilistic projections. We validate the method using a suite of one-at-a-time cross-validation experiments involving Atlantic meridional overturning circulation (AMOC), its temperature-based index, as well as Korean summer mean maximum temperature. In these experiments the method tends to exhibit superior skill over a trend-only Bayesian model averaging weighting method in terms of weight assignment and probabilistic forecasts. Specifically, mean credible interval width, and mean absolute error of the projections tend to improve. We apply the method to a problem of projecting summer mean maximum temperature change over Korea by the end of the 21$^{st}$ century using a multi-model ensemble. Compared to the trend-only method, the new method appreciably sharpens the probability distribution function (pdf) and increases future most likely, median, and mean warming in Korea. The method is flexible, with a potential to improve forecasts in geosciences and other fields.

# 1 Introduction

A common forecasting problem is one of probabilistic multi-model forecasts of a stochastic dynamical system [1–18]. Sometimes, when a collection of complex dynamical models is used to provide multi-model forecasts, these forecasts are weighted according to model performance compared to observations [1,5,10,19–23]. The Bayesian approach to this problem assumes that associated with $k$ dynamical models are $k$ competing statistical models $M_i$ for vector of observations *y*. These statistical models result in a conditional probability density function (pdf) for *y* given that $M_i$ is reasonable, $p(y|M_i)$. Typically, in a multi-model evaluation context, the pdf $p(y|M_i)$ is a multivariate statistical distribution centered on *i*th dynamical model trend $x_i$. Each model is associated with a prior belief in its adequacy ("prior") $p(M_i)$, which can be derived from previous work, or may be more subjective. The posterior probability, or weight, for each model *i* given the observations is then found using Bayes theorem [24]:

$$p(M_i|y) \propto p(y|M_i)p(M_i) \qquad (1)$$

Specifically, the posterior probability of each statistical (and corresponding dynamical) model is the likelihood of observations *y* coming from the model (given by the pdf $p(y|M_i)$), multiplied by the model prior.

In ensemble modelling, models are usually judged on how well they represent the mean state of the system, its trend, or spatio-temporal fields [1,3,6,14,22,23]. However, it is increasingly being recognized that variability is of utmost importance for future prediction. Specifically, for some systems (stochastic dynamical systems) the stationary pdf of the equilibrium solution is directly affected by system dynamics (i.e., the nonlinear operator in the ordinary differential equations) through the Fokker-Planck (Kolmogorov forward) equation. Recent climate science work identifies variability as a key factor impacting climate projections [6,25]. Furthermore, variability has been used as a novel and effective constraint for climate sensitivity [26]. In addition, variability also has major relevance for forewarning of critical thresholds (i.e., a forcing value above which the underlying system shifts to a new equilibrium; [27]). Specifically, an increase in variance or lag-1 autocorrelation with time, as well as skewness and kurtosis, have been used as such early warning indicators [28–31]. This motivates using variability properties of the system as a novel metric to assess performance of multiple dynamical system models.

Several new studies break important new ground by incorporating variability into the weighting [17,32–34], but they typically assume stationarity of the pdf of the system [17,32,33], or cannot work with complex dynamical models [34]. Some previous work does explicitly weight dynamical models by performance in variability and trends in a statistically-sound way [35]. However, the method in its current form works only for linear trends (as a function of time) and does not account for autocorrelation in the variability.

Here we propose a novel method to weight models of complex dynamical systems by their performance in autocorrelation, variability, and a potentially nonlinear trend (i.e., nonlinear with time) compared to observations, and to make probabilistic forecasts. The method is based on Bayesian Model Averaging (BMA) [20,21]. While the framework is Bayesian, it deviates from traditional Bayesian theory in some steps of the estimation process. We highlight these deviations where they arise in more detail in later sections. Consequently, we call our approach "quasi-Bayesian". Using several simulated and observed datasets (involving AMOC, its temperature-based index, and summer mean maximum temperature over Korea) we show that the new method results in better weighting and tends to improve forecasts of system mean change under new conditions compared to when trend-only BMA weighting is used. Thus, this work has implications for improving projections of many environmental systems. The approach is not restricted to linear trends, making it relatively easy to apply to new datasets. Finally, we apply the method to a real case problem of projecting future summer mean temperature changes over Korea.

The rest of the paper is structured as follows. Section 2 describes the novel methodology to weight models by trend and variability performance, to combine those weights, to make multi-model weighted projections, as well as the computational details. The main interest here is not the procedure for obtaining the trend and variability components, but the algorithm for model weighting. In Section 3 we describe leave-one-out cross-validation experiments to test method performance against a trend-only BMA method. Here we also provide the specific details on how the trend and variability components were extracted from the data. Section 4 describes the results of these experiments. Section 5 discusses the application of the method to make multi-model probabilistic projections of Korean summer mean maximum temperature change. Section 6 briefly discusses the main findings of the study and places it in context of prior work. Section 7 discusses the limitations of the work, and Section 8 presents conclusions.

# 2 Materials and Methods

## 2.1 Overview of the method

At the start of the analysis, we assume that we have a collection of dynamical model time series outputs, and that these outputs can be decomposed into long-term trend and variability components. The details of this decomposition are not critical for this study, as we focus on the statistical methodology for the weighting. The weights (or probabilities) for the two submodels are calculated separately, using the Bayesian statistical paradigm, and then combined. The combined weights can then be used to make predictions (Fig 1).

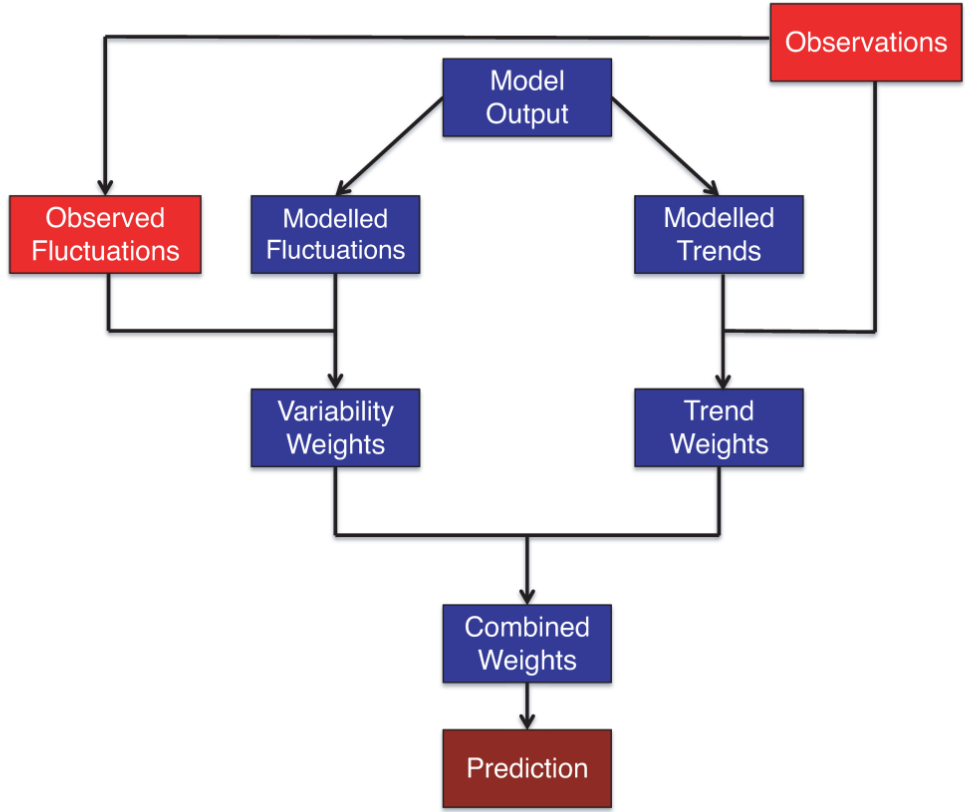

**Fig 1.** Schematic illustrating the proposed "trend+var" method.

## 2.2 Notation and decomposition of model output.

Consider that *k* models are available. We postulate that each dynamical model is associated with a statistical model $M_i$ for the observations. $M_i$ can be thought of as a statistical event, which when true indicates that *i*th dynamical model is a reasonable representation of real system. $M_i$ consists of two submodels: a trend submodel $M_{T,i}$ (related to the trend in the system), and a variability submodel $M_{V,i}$ (modelling internal fluctuations in the system). When $M_{T,i}$ is true, the *i*th dynamical model correctly captures the trend of the system. Likewise, when $M_{V,i}$ is true, the *i*th dynamical model correctly captures the variability of the system. Alternatively, we can consider the model for anomalies scaled by the mean ($M_{V0,i}$). Each model produces time series output of a physical quantity during the period when observations are available ("calibration period"), as well as under new forcing conditions, usually associated with future system projections ("projection period"). We are interested in finding the probability distribution of a change of the system mean $\Delta$ between a "projection reference period" (typically the same as the calibration period) and the projection period. We denote the raw calibration period model output from the $i$th dynamical model by vector $x'_i = (x'_{i,1}, ..., x'_{i,n})$ where superscript "'" indicates that the output is raw (un-smoothed), and $n$ is the length of the record. The model output is a regularly spaced time series. We consider decomposition of the form:

$$x'_i = \underbrace{x_i}_{\text{trend}} + \underbrace{\Delta x_i}_{\text{anomalies}} \qquad (2)$$

We will use the term "anomalies" to refer to the variability component of the time series. The trend $x_i$ can be either a linear trend, or a more flexible nonlinear trend obtained, for example, from robust locally weighted regression [36]. We assume that this decomposition is deterministic, unique, and is performed before the start of the main analysis. We also assume that the estimate of the trend is a reasonable proxy for the true unknown trend. While it may be possible to also incorporate the uncertainty in this decomposition, we leave it to future work. The focus here is not on how to properly decompose a time series into a long-term trend and variability, but on the novel methodology for weighting by performance in both. See [18] for an example of use of an alternative methodology to decompose the data. The use of alternative methods for data decomposition is subject of future research. We describe the decomposition method we use for each dataset in Section 3. The same decomposition is also applied to the observed time series $y'$:

$$y' = y + \Delta y. \qquad (3)$$

Another option is relative decomposition. It takes the following form:

$$x'_i = \underbrace{x_i}_{\text{trend}} + \underbrace{\bar{x}_i \Delta x^0_i}_{\text{anomalies}},  \qquad (4)$$

where $\bar{x}_i$ is the deterministic sample mean of the $i$th dynamical model output, and $\Delta x^0_i$ are normalized anomalies; and similarly for the observations:

$$y' = y + \bar{y}\Delta y^0_i, \qquad (5)$$

where $\bar{y}$ is the observed mean.

Next subsections contain the following: subsection 2.3 discusses the trend submodel weighting (which largely follows previous work), subsection 2.4 centers on the variability submodel weighting, section 2.5 discusses combining the component weights for each model, section 2.6 is dedicated to procedure for making weighted multi-model projections, and section 2.7 presents computational details on the implementation of the method.

## 2.3 Weighting the Trend Submodels

The trend submodel weighting is implemented following prior work, and full details are provided there [9]. Essentially, this method is BMA that also considers the uncertainty due to model error, and uncertainty in statistical properties of data-model residuals. Here, we consider $k$ competing statistical models $M_{T,i}$ for raw observations $y'$. We stress that statistical and dynamical models are conceptually related: i.e., if the statistical model $M_{T,i}$ is true, it implies that the associated $i$th dynamical model correctly represents the trend in the system. Each $M_{T,i}$ is a hierarchical statistical model that connects modelled deterministic trend from the $i$th model during the calibration period $x_i$ to real system trend $y$, and then the system trend to actual observations $y'$ (Eq. 6):

$$\begin{cases} y = x_i + f\varepsilon_D \\ y' = y + \varepsilon_{NV}, \end{cases} \qquad (6)$$

where $f\varepsilon_D$ is random discrepancy (long-term model error), and $\varepsilon_{NV}$ is random internal variability (as well as short-term observational error).

Here we deviate somewhat from orthodox Bayesian practice. A typical Bayesian approach would assume a distributional form for the discrepancy vector $f\varepsilon_D$. However, because this error is likely long-term dependent, and the probability distributions for its components are not necessarily normal, finding and justifying a proper parametric model for it is non-trivial. To deal with this conundrum, we adopt an approach inspired by prior work [37]. We postulate that model error can be derived from inter-model trend differences. The reasoning for this implementation is as follows. Imagine a particular trend submodel $M_{T,i}$ represents the "true" system. Associated with this system is trend $x_i$ and pseudo-observations $x'_i$. If only the rest of the models are available to the researcher, then the best-fit model $j$ to these pseudo-observations is associated with trend $x_j$. The difference between the best model and the pseudo-observed trends is then the unscaled error of the $j$th model. Thus, we obtain samples for unscaled discrepancy $\varepsilon_D$ directly from the differences between each model's trend and the next-closest model trend (see [9] for details). We acknowledge that this parameterization is simplified; model error is an emergent research topic [37]. We thus hope this work can galvanize more research on parametrizing model error.

The second non-orthodox idea, is related to the deterministic $f$ factor ("error expansion factor", Eq. 6). This factor is a new addition to the model presented in previous work [9]. $f$ is a parameter that scales $\varepsilon_D$ to account for potential overconfidence. The non-orthodox idea relates to the procedure for selecting $f$. Specifically, we do not estimate $f$ from present-day observations as a strict Bayesian would do, but rather we select *f* that results in correct coverage of the 90% posterior credible intervals during cross-validation experiments (different *f* for each dataset). The reason for this is as follows. Using just present-day observations to estimate *f* may produce small *f* values that result in overconfident future projections. This is because models have been developed so that they match observed data. Philosophically, present-day model-data agreement may be due to overfitting, and may not be reflective of the actual amount of error in the models.

The internal variability $\varepsilon_{NV}$ (Eq. 6) is modelled as an AR(1) process with random parameters $\boldsymbol{\theta} = (\sigma, \rho)$, where $\sigma$ is innovation standard deviation and $\rho$ is autocorrelation. Following Bayes theorem, and marginalization theorem, the trend model weights are then calculated as:

$$p(M_{T,i}|\boldsymbol{y}') \propto p(M_{T,i}) \iint p(\boldsymbol{y}'|\boldsymbol{y}, \boldsymbol{\theta}, M_{T,i}) p(\boldsymbol{\theta}) p(\boldsymbol{y}|M_{T,i}) d\boldsymbol{y} d\boldsymbol{\theta}. \qquad (7)$$

Here, $p(M_{T,i})$ denotes the prior for the *i*th trend model, $p(\boldsymbol{y}'|\boldsymbol{y}, \boldsymbol{\theta}, M_{T,i})$ is the AR1 likelihood resulting from the bottom line of Eq. (6), $p(\boldsymbol{\theta})$ denotes the prior for the AR1 parameters, and $p(\boldsymbol{y}|M_{T,i})$ is obtained according to the top line of Eq. (6) using samples from $f\varepsilon_D$ as discussed above. Unlike the previous work [9], here we assume uniform prior probabilities for trend models $p(M_{T,i})$. The integral is evaluated using Monte Carlo integration, which is simpler to implement than Markov chain Monte Carlo methods used in some studies [2]. For the relative low dimension parameter space that we deal with here, simple Monte Carlo is adequate. Additional experiments suggest the sample size we use for the Monte Carlo integration is reasonable to minimize Monte Carlo error (Text A in S1 File). Once calculated, the weights are normalized to sum to 1 to facilitate interpretation as probabilities. We provide technical details in Text A in S1 File.

## 2.4 Weighting the Variability Submodels

Variability models are weighted using similar ideas to the ones used in trend weight estimation. We consider $k$ competing statistical models for calibration period anomalies observations $\Delta \boldsymbol{y} = (\Delta y_1, \Delta y_2, \ldots, \Delta y_n)$ (see Eq. (3)). Each $i$th variability model $M_{V,i}$ models the anomalies hierarchically in the following form:

$$\begin{cases} \boldsymbol{\theta}_y^V = \breve{\boldsymbol{\theta}}_{M,i}^V + f\boldsymbol{\varepsilon}^V \quad (9) \\ \Delta y_t = \rho_y \Delta y_{t-1} + w_t, \end{cases} \qquad (8)$$

where $\boldsymbol{\theta}_y^V = (\sigma_y, \rho_y)$, are autocorrelation and innovation standard deviation of the real climate, $\breve{\boldsymbol{\theta}}_{M,i}^V = (\breve{\sigma}_{M,i}, \breve{\rho}_{M,i})$ are summary statistics of autocorrelation and innovation standard deviation from $i$th model anomalies, $f\boldsymbol{\varepsilon}^V$ is model error (where $\boldsymbol{\varepsilon}^V = (\varepsilon_\sigma, \varepsilon_\rho)$, and $f$ is a deterministic scaling factor to widen the distribution to correct for potential overconfidence), and $w_t \sim N(0, \sigma_y^2)$. The top line of Eq. (8) connects real system anomaly properties to model summary statistics, and the bottom line shows that observed anomalies are modelled as red noise with parameters $(\sigma_y, \rho_y)$ of the real system.

Thus, in the top line of Eq. (8) instead of performing full posterior sampling to obtain samples for real system autocorrelation and innovation standard deviation parameters $\boldsymbol{\theta}_y^V$ we assume they are centered around summary statistics $\breve{\boldsymbol{\theta}}_{M,i}^V$ of $i$th physical model anomalies with an additive error $f\boldsymbol{\varepsilon}^V$. Each model's summary statistics are taken as the corresponding MLE estimates. Again, we refrain from assuming any parametric form for $\boldsymbol{\varepsilon}^V$. Similar to the error for the trend model, here we also assume samples for $\boldsymbol{\varepsilon}^V$ are obtained from differences between each model MLE summary statistics $\breve{\boldsymbol{\theta}}_{M,i}^V = (\breve{\sigma}_{M,i}, \breve{\rho}_{M,i})$ and the next-closest model summary statistics $\breve{\boldsymbol{\theta}}_{M,j}^V = (\breve{\sigma}_{M,j}, \breve{\rho}_{M,j})$. The next-closest model is found as follows: for each model *i* we compare the conditional likelihood of *i*th model anomalies given AR(1) parameters of other variability submodels $p(\Delta x_i | \breve{\boldsymbol{\theta}}_{M,j}^V)$, $j \neq i$ under the AR(1) statistical model, and find a model *j* that maximizes this likelihood. We also add a sample of zero vector (0,0) to $\boldsymbol{\varepsilon}^V$ for computational stability. We post-multiply these samples by a scaling factor *f* to obtain samples for $f\boldsymbol{\varepsilon}^V$. *f* is the same parameter that is used to scale trend model discrepancy (Section 2.3).

This approach gives us only *k*+1 samples from $f\boldsymbol{\varepsilon}^V$. To obtain a larger number of samples which are well-dispersed, we add to $f\boldsymbol{\varepsilon}^V$ realizations from an independent bivariate normal distribution with standard deviations in each dimension set to 1/5 of the original *k*+1 sample ranges. We use the value of 1/5 because it results in samples with a reasonably smooth density that preserves large scale cross-correlation structure between the original *k*+1 samples of $\varepsilon_\sigma$ and $\varepsilon_\rho$, and provides a decent approximation to the underlying pdf for $f\boldsymbol{\varepsilon}^V$ (Fig A in S1 File). Sensitivity tests indicate that using lower standard deviations can degrade the smoothness of the pdf (not shown).

Then, the posterior probability of the variability model *i* is, using Bayes rule [24] and probability rules:

$$p(M_{V,i}|\Delta y) = \int p(M_{V,i}, \boldsymbol{\theta}_y^V|\Delta y)\, d\boldsymbol{\theta}_y^V \propto p(M_{V,i}) \int p(\Delta y|M_{V,i}, \boldsymbol{\theta}_y^V) p(\boldsymbol{\theta}_y^V|M_{V,i})\, d\boldsymbol{\theta}_y^V, \qquad (9)$$

where $p(\Delta y|M_{V,i}, \boldsymbol{\theta}_y^V)$ is an AR1 likelihood function, $p(\boldsymbol{\theta}_y^V|M_{V,i})$ is sampled using the top line of Eq. (8) using bootstrapping from $f\boldsymbol{\varepsilon}^V$ as described above, and $p(M_{V,i})$ is the prior probability ("prior") for the *i*th variability submodel. We assume equal priors for all submodels. This integral is also evaluated using Monte Carlo integration. Specifically, we sample from the conditional pdf of real system summary statistics given each variability model $p(\boldsymbol{\theta}_y^V|M_{V,i})$ as described above, and for each sample we calculate the conditional likelihood for the observed anomalies $p(\boldsymbol{\Delta y}|M_{V,i}, \boldsymbol{\theta}_y^V)$. The integral is approximated as a simple mean of the conditional likelihoods across the samples. Probabilities are calculated for each submodel and are normalized to sum up to 1. The implementation using relative variability $M_{V0}$ is identical except the residuals $\boldsymbol{\Delta x_i}$ and $\boldsymbol{\Delta y}$ are normalized by the respective model and observational means prior to the analysis. We provide technical details on the implementation in Text B in S1 File.

## 2.5 Combined Weights and Bayesian Model Averaging

In the next step, the weights for the two submodels are put together to form a single combined model weight. Using probability laws:

$$p(M_i|\boldsymbol{y}, \boldsymbol{\Delta y}) = p(M_{T,i}, M_{V,i}|\boldsymbol{y}, \boldsymbol{\Delta y}) = p(M_{T,i}|M_{V,i}, \boldsymbol{y}, \boldsymbol{\Delta y}) \times p(M_{V,i}|\boldsymbol{y}, \boldsymbol{\Delta y}). \tag{10}$$

We make two simplifying assumptions. First, we observe that in the datasets described in Section 3 typically the relationships between the variability summary statistics $\breve{\sigma}_{M,i}$ and $\breve{\rho}_{M,i}$ on one hand, and trend model probability on the other hand, appear to be weak (Figs B-K in S1 File). In addition, the corresponding linear coefficients are almost always weak (weak is defined as the absolute values less than 0.5). Assuming that the relationships based on the sample summary statistics are a good proxy for those based on the population properties, we make an assumption that the probability of the trend model is independent of the variability model:

$$p(M_{T,i}|M_{V,i}, \boldsymbol{y}, \boldsymbol{\Delta y}) \approx p(M_{T,i}|\boldsymbol{y}, \boldsymbol{\Delta y}) = p(M_{T,i}|\boldsymbol{y}'), \tag{11}$$

which allows us to directly plug in trend model weights obtained using the method in Section 2.3. Second, since only anomalies are used to weight the variability model:

$$p(M_{V,i}|\mathbf{y}, \mathbf{\Delta y}) = p(M_{V,i}|\mathbf{\Delta y}). \tag{12}$$

This quantity is obtained following Section 2.4. As a result, the combined weights can be expressed as a product of the trend and variability submodel weights:

$$p(M_i|\mathbf{y}, \mathbf{\Delta y}) = p(M_{T,i}|\mathbf{y}, \mathbf{\Delta y}) \times p(M_{V,i}|\mathbf{\Delta y}). \tag{13}$$

We stress that even though the independence assumption generally appears reasonable here, it may not always apply. Hence, it is recommended to check it when applying the methodology to new datasets. Incorporating the potential dependence between the trend and variability submodels into our framework is the subject of future research. Once calculated, the probabilities are normalized to sum up to 1, meaning that we restrict our probability space to the union of available models $M_i$.

## 2.6 Future Projections

Future model projections are implemented largely following previous work [9]. Once the weights are obtained, the statistical model for system change between projection reference and projection periods $\Delta$ follows the BMA formula [20,21]:

$$p(\Delta|D) = \sum_{i=1}^{k} p(\Delta|M_i, D) p(M_i|D) = \sum_{i=1}^{k} w_i p(\Delta|M_i, D), \tag{14}$$

where $D = (\mathbf{y}, \mathbf{\Delta y})$ is collection of all available observations, $p(\Delta|M_i, D)$ is conditional probability for the change given than $i$th dynamical model is correct, and $w_i = p(M_i|D)$ is the probability for the $i$th model (i.e., model weight) found earlier (Eq. (13)) as the product of the trend and variability model probabilities. This represents a skill-weighted mixture of pdfs from individual models. Here we consider $\Delta$ to be a simple difference between projection period mean and forecast reference period mean. Future predictions are largely modelled following prior work [9]. Just as for the calibration period, we assume a deterministic decomposition of projection period output into trend and anomalies:

$$x_i'^{(f)} = \underbrace{x_i^{(f)}}_{\text{trend}} + \underbrace{\Delta x_i^{(f)}}_{\text{anomalies}} \qquad (15)$$

The exact decomposition method for each dataset is listed in Section 3. Next, we consider the following statistical model for dynamical system time-series projections (all quantities are vectors):

$$y'^{(f)} = x_i^{(f)} + b^{(f)} + \varepsilon_{S,i}^{(f)}, \qquad (16)$$

where $y'^{(f)}$ is the projection time series, $x_i^{(f)}$ is $i$th model trend output from Eq. (15), $\boldsymbol{b}^{(f)} = b^{(f)}\mathbf{1}$ is random time-constant bias, and $\varepsilon_{S,i}^{(f)}$ is random short-term internal variability in each model. Thus, we assume that if $i$th model is correct, the vector projection is the sum of $i$th model trend, a time constant bias, and internal variability. Here we again deviate somewhat from the traditional Bayesian theory in that the components of this model are partially informed by inter-model differences, and by model output during cross-validation experiments. Such steps are necessitated by the absence of actual system observations over the projection period to inform us about these components. We model the bias parameter as $b^{(f)} \sim N(0, f\breve{\sigma}_b^{(f)})$ where $\breve{\sigma}_b^{(f)}$ is sample standard deviation of future period-mean next-closest model differences (where next-best is used in the $l_1$ distance sense), and $f$ is the deterministic model error expansion factor (the same factor that is used for model weighting). Two different formulations are implemented for internal variability. In the first formulation ("boot"; [9]) we use simple bootstrapping from $\Delta x_i^{(f)}$ to generate internal variability samples. In the alternative formulation ("ar1") we sample $\varepsilon_{S,i}^{(f)}$ as a red noise process with parameters $\boldsymbol{\theta}_i^{(f)} = (\breve{\sigma}_i^{(f)}, \breve{\rho}_i^{(f)})$, the sample innovation standard deviation and autocorrelation of future anomalies. An improvement would be to consider the uncertainty in the AR1 parameters; we do not do this here to simplify the method. To obtain projection period mean changes from the reference period, we take weighted samples of future projections using Eq. (14) and (16), and simply subtract projection reference period mean modeled value for each model. As in previous work [9], we use 100,000 samples for all experiments.

The overall algorithm for the method is illustrated in Fig 2. The method estimates model weights from calibration period observations, and has one fixed parameter $f$, quantifying model error. Larger $f$ values lead to higher model errors, and as a result broader projections with higher coverage of the 90% posterior credible intervals. Unlike standard Bayesian analysis, we first choose $f$ to obtain approximately correct empirical coverage of the 90% posterior credible intervals during cross-validation. For the cross-validation, each model is selected as the "truth" one-at-a-time. Models are weighted using the output from the "true" model. The "true" model is then excluded from the model set, and the future weighted projections from the remaining models are compared to the output from the "true" model. Once $f$ achieves approximately correct empirical coverage, the method is used for actual projections constrained by real observations. If there are many replicates (or regions) of the system, cross-validation can also be performed by splitting the calibration period into two subperiods. In step 1, observations during the first subperiod in each region/replicate can be used to assign replicate/region-specific weights. In step 2, observations during the second subperiod can test the empirical coverage of the posterior credible intervals. Here, however, we focus on the one-at-a-time cross-validation using future model output. This is because (i) the length of historical record for which high-quality observations are available is too short for most of the experiments [38,39], (ii) observational records suffer from observational

"trend+var" ALGORITHM

INPUT: trend and fluctuations model output for each model $i$ for calibration period $x_i$ and $\Delta x_i$, variability summary statistics $\breve{\theta}^V_{M,i}$, corresponding trend and fluctuations observations $y$ and $\Delta y$, projection period model output $x_i^{(f)}$ and $\Delta x_i^{(f)}$, model means during projection reference period $\bar{x}_i^{(r)}$.
OUTPUT: probability distribution of a metric change $\Delta$ between the projection and the projection reference periods

**while** coverage of 90% posterior credible interval $\approx$ 90% **do**
  set $f$ #Choose f by trial and error
  **for** $i$ in 1:$k$ **do** #Perform one-at-a-time cross-validation; select each model as "truth"
      assign $y^i$ to $x_i$
      assign $\Delta y^i$ to $\Delta x_i$
      **for** $j$=1:$k$ **do**
        find $p(M_{V,j}|\Delta y^i)$ using Eq. 9
        find $p(M_{T,j}|y'^i)$ using Eq. 7
        find $p(M_j|y^i,\Delta y^i)$ using Eq. 13
      **end do**
      find $p(\Delta|D^i) = \sum_{j \neq i} p(\Delta|M_j, D^i) p(M_j|D^i)$ #where $D^i = (y^i, \Delta y^i)$
      **compare** the 90% credible interval of $\Delta$ to $\Delta_i = \bar{x}_i^{(f)} - \bar{x}_i^{(r)}$
  **end do**
**end do**
# Use calibrated $f$ from the cross-validation, perform real projections
**for** $i$=1:$k$ **do**
  find $p(M_{V,i}|\Delta y)$ using Eq. 9
  find $p(M_{T,i}|y')$ using Eq. 7
  find $p(M_i|y, \Delta y)$ using Eq. 13
**end do**
find $p(\Delta|D) = \sum_{i=1}^{k} p(\Delta|M_i, D) p(M_i|D)$

**Fig 2:** "trend+var" algorithm.

errors, and (iii) climate signal (e.g., the magnitude of climate changes) is quite low in the historical period. We choose various variables and periods to test the method under different conditions.

## 2.7. Computational details

All experiments have been performed on an Intel Xeon CPU X5650 @ 2.67GHz GNU/Linux 2.6.18-164.el5 supercomputer, using R programming language version 3.3.3. For other required packages the following versions were used: mblm 0.12 and KernSmooth 2.23-15. We provide the R code as a supplementary file File S2. This code is provided under the GNU general public license v3.

In the next section we describe several cross-validation experiments for our method and compare the performance of the method (which we call hereafter "trend+var") with a BMA method where all variability submodel weights are set to equal (termed hereafter "trend"). Note that "trend" method is BMA which still weights models by their performance in terms of trend.

# 3 Leave-One-Out Cross-Validation Experiments to Test Method Skill

## 3.1 Overview of Leave-One-Out Cross-Validation Experiments

To evaluate method performance, we carry out leave-one-out cross-validation experiments with several simulated and observed datasets: (i) Atlantic meridional overturning circulation (AMOC) strength [Sv] from 13 global climate models (GCMs) (AMOC experiment), (ii) Korean summer mean maximum temperatures from 29 GCMs (Korea_temp), (iii) Korean temperatures with an extended calibration period (Korea_temp_long), (iv) winter East Sea surface temperatures (SSTs) (Winter SST Experiment), (v) temperature-based AMOC Index (temperature in northern North Atlantic "gyre" minus Northern Hemisphere temperature) from 13 GCMs (AMOCIndex), and (vi) the same as (v) but also considering information from climate observations (AMOCIndex_obs). We discuss each experiment in greater detail in the following subsections. The cases differ in terms of the calibration, projection, and projection reference periods (Table 1). In experiments involving model output only, each of the models is selected as "truth" one at time, and its output is used to weight the models. Then, during the validation period, the projected pdfs of changes using the remaining models are compared to the "true" model output. The set-up for the AMOCIndex_obs is slightly different: both calibration and validation periods have available instrumental observations. Here, instead of selecting each model output as pseudo-observations one-at-a-time, we simply use actual observations to both weight the climate models, and to evaluate the projections. All experiments are performed with both "trend" and "trend+var" methods. Both methods have been calibrated for each experiment to have approximately correct coverage (correct % of cases where the "truth" is outside the 90% posterior credible intervals) by adjusting the model error expansion factor $f$ (Table 1). The calibrated values of $f$ for the AMOCIndex experiments are also used for the corresponding AMOCIndex_obs experiments. We focus on the Winter_SST experiment here, however summary results for all experiments are also provided.

**Table 1.** Basic information about the design of leave-one-out cross-validation experiments, and the method performance. **Bold font** indicates improvement of the "trend+var" method, compared to the "trend" method. $k$ is the number of models in the ensemble; MCIW is mean 90% credible interval width; MAE is mean absolute bias of the mean; CIW is 90% credible interval width; AB is absolute bias of the mean.

| Experiment | k | Calibration Period | Projection Reference Period | Projection Period | Trend f | Trend+ Var f | Metric | Trend | Trend+ Var |
|---|---|---|---|---|---|---|---|---|---|
| AMOC | 13 | 1880-2004 | 1960-1999 | 2060-2099 | 1.5 | 1.5 | MCIW | 9.53 Sv | **9.46 Sv** |
|  |  |  |  |  |  |  | MAB | 2.30 Sv | **2.06 Sv** |
| Korea_temp | 29 | 1973-2005 | 1973-2005 | 2081-2100 | 1.55 | 0.75 | MCIW | 4.34 K | **3.23 K** |
|  |  |  |  |  |  |  | MAB | 1.01 K | **0.88 K** |
| Korea_temp_long | 29 | 1950-2005 | 1950-2005 | 2081-2100 | 2.22 | 2.3 | MCIW | 4.28 K | **3.60 K** |
|  |  |  |  |  |  |  | MAB | 0.96 K | **0.79 K** |
| Winter_SST | 26 | 1941-2000 | 1941-2000 | 2061-2000 | 2.5 | 2.05 | MCIW | 3.40 K | **2.90 K** |
|  |  |  |  |  |  |  | MAB | 0.86 K | **0.78 K** |
| AMOCIndex | 13 | 1880-1945 | 1880-1945 | 1965-2004 | 3.75 | 3.75 | MCIW | 1.42 K | 1.44 K |
|  |  |  |  |  |  |  | MAB | 0.23 K | 0.23 K |
| AMOCIndex_obs | 13 | 1880-1945 | 1880-1945 | 1965-2004 | 3.75 | 3.75 | CIW | 1.42 K | 1.43 K |
|  |  |  |  |  |  |  | AB | 0.42 K | **0.41 K** |

## 3.2 AMOC Experiment

For the AMOC experiment (Table 1), data extraction and processing largely follow previous work [9]. The Climate Model Intercomparison Project phase 5 (CMIP5; [40]) model output for this (and other) experiments has been obtained from the ESGF LLNL portal [41]. Future forecasts use the RCP8.5 emissions scenario [42]. We use robust locally-weighted "lowess" regression [36] to obtain the trend model component during the calibration period, and Theil-Sen slopes [43] – in the validation period. We set the "lowess" smoother span parameter to 0.8 during the smoothing. We use this span value because it appears effective at removing interdecadal variability. The smoothed model output is illustrated in Fig 3. Importantly, we see nonlinearities in the modeled trends. Previous variability weighting work does not account for such nonlinearities [35]. During the trend weighting we use smoothed output as anomalies with respect to the entire calibration period. We use normalized (by the absolute AMOC) anomalies to weight the variability models. Future projections use the "boot" variant of the method.

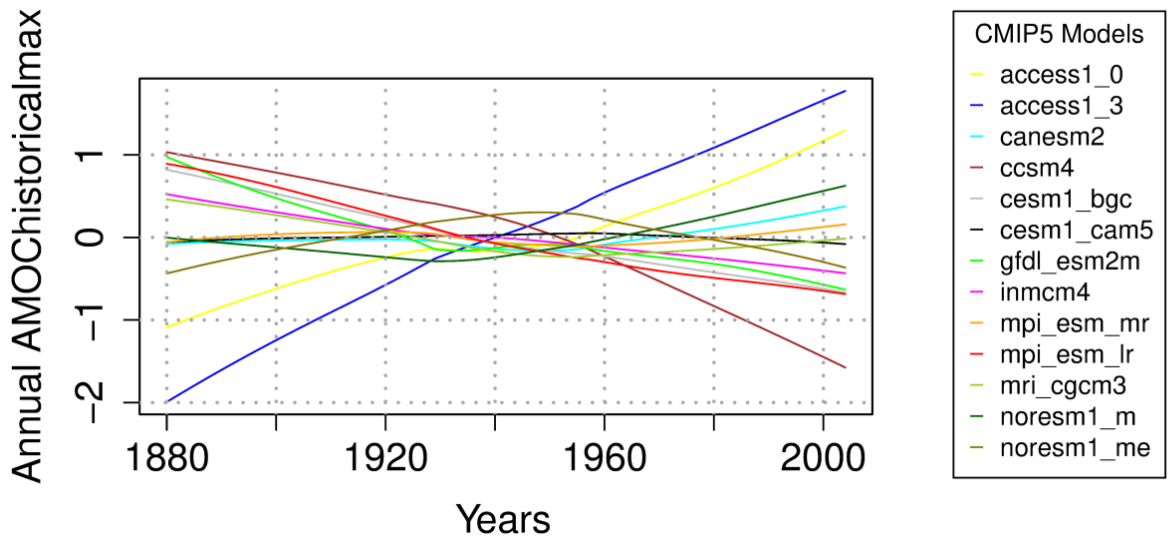

**Fig 3:** AMOC anomaly trends for the calibration period [Sv], as simulated by the CMIP5 climate models.

### 3.3 Korea_temp and Korea_temp_long Experiments

Korea_temp and Korea_temp_long differ only in the calibration periods and the error expansion factors *f*, with Korea_temp_long using a longer calibration period. These experiments use output from historical and future RCP8.5 runs of 29 CMIP5 model runs (Table 1, Table A in S1 File). First, Korean daily maximum temperatures are calculated as spatial averages over land grid cells (cells with more than 80% land) between 34–40ºN and 125–130ºE [38]. The JJA (June,

July, August) means are then obtained for each year. Theil-Sen slopes are used for obtaining model trends during the model output decomposition. During the weighting, smoothed output is used as anomalies with respect to the entire calibration period. Future projections use the "boot" variant of the method. Note that the Korea_temp "trend+var" experiment has a slightly elevated coverage of 93%. Decreasing $f$ to obtain approximately 90% coverage is expected to improve performance metrics, but also to make probability densities too discontinuous. Hence, we use the value of $f=0.75$.

## 3.4 Winter_SST Experiment

Winter_SST experiment uses winter sea surface temperatures from the East Sea from historical and future RCP8.5 runs of 26 CMIP5 climate models (Table 1, Table B in S1 File). We select this dataset because we find considerable relationships between present-day internal variability properties and future SST change in this region and season (Fig 4; for model number corresponding to each model see Fig L in S1 File). We define the East Sea as the area between 35 °N and 42°N, and between 130 °E and 139 °E. We use a simple average of all ocean points in this region. During the weighting we use the output as anomalies with respect to the calibration period. Furthermore, we use Theil-Sen slopes to obtain model output trends. Future projections use the "ar1" variant of the method, since we detect a considerable autocorrelation in the model output anomalies.

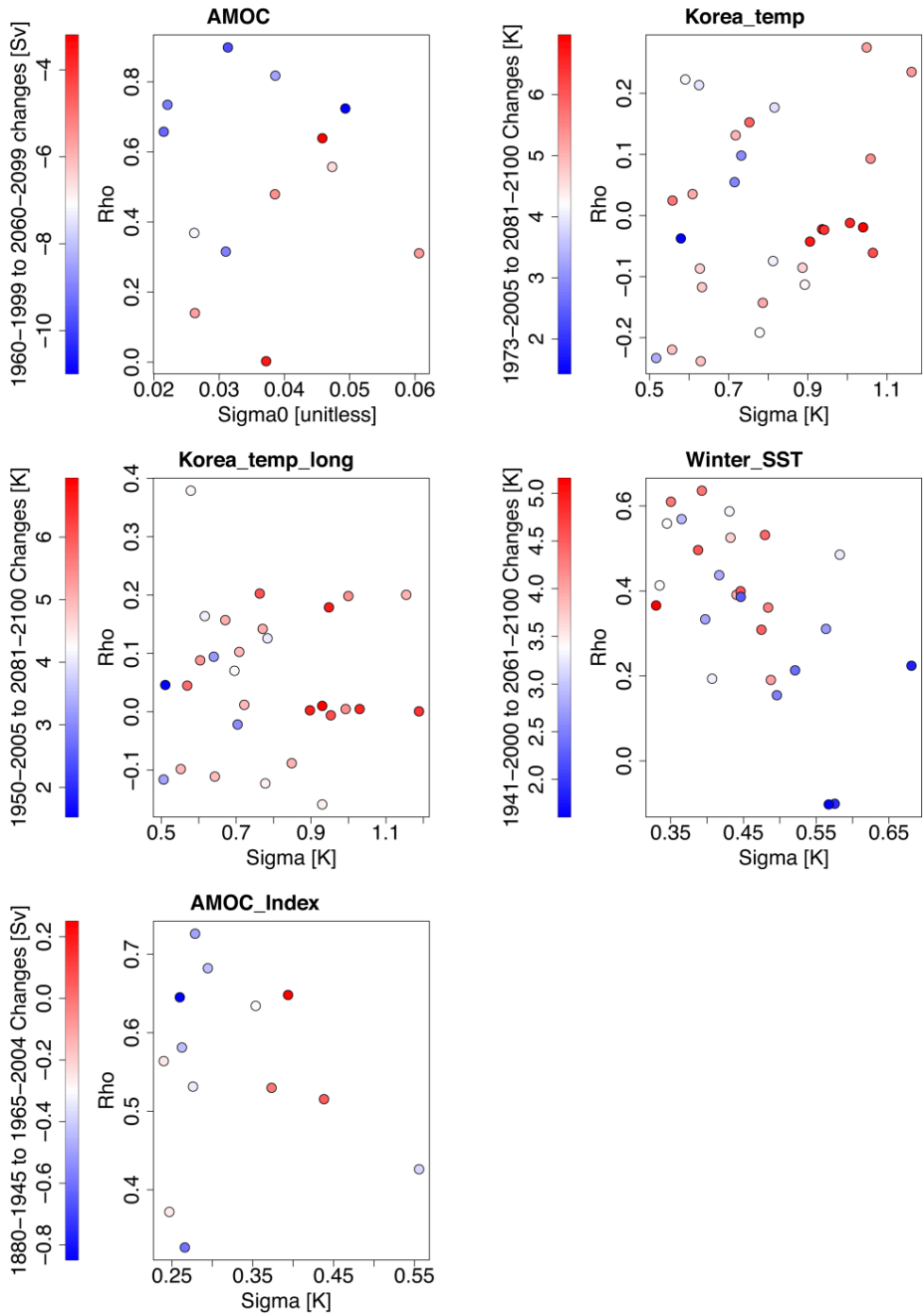

**Fig 4.** Relationship between sample innovation standard deviations $\breve{\sigma}_{M,i}$ (normalized in the AMOC experiment) and sample lag-1 autocorrelations of model anomalies during the calibration $\breve{\rho}_{M,i}$ for each model, and projected changes (projected mean minus reference period mean) for the experiments using simulated data.

## 3.5 AMOCIndex Experiment

AMOCIndex experiment (Table 1) relies on historical output from the same 13 CMIP5 models used for the AMOC experiment. AMOC Index is defined as sea surface temperature in northern North Atlantic "gyre" minus Northern Hemisphere temperature. It is physically linked to northward heat transport by the AMOC, and hence can be used as a proxy for AMOC [9,44]. Data extraction and processing follow [9], with a few changes. The Index is used as an anomaly with respect to the entire historical period 1880-2004. We then use a portion of the historical period (1880-1945) for calibration, and another portion (1965-2004) for projections. Smoothing is performed using Theil-Sen slopes. Projections use the "ar1" variant of the method.

## 3.6 AMOCIndex_obs Experiment

For the AMOCIndex_obs we use actual observations both to weight the models, and to validate the probabilistic projections. Otherwise, the experiment relies on the same model output as AMOCIndex experiment. The observations are a simple average of two AMOC Index versions: one calculated with ERSSTv4 SSTs [45–47], and one with COBE-SST2 SSTs [48]. ERSSTv4 data is publicly curated by National Oceanic and Atmospheric Administration [49], while COBE-SST2 observations are provided by M. Ishii on the servers of Hokkaido University, Japan [50]. Both versions use GISTEMP Northern Hemisphere temperatures [51]. GISTEMP observations are maintained by NASA Goddard Institute for Space Studies [52]. For comparison with model output the COBE-SST2 SSTs are first interpolated to a 2×2° grid using bilinear interpolation, while the ERSSTv4 observations are already on such a grid. For both "trend" and "trend+var" experiments, $f$ is taken from corresponding AMOCIndex experiments.

# 4 Results of Leave-One-Out Cross-Validation Experiments

The new method tends to be better able to correctly identify the "true" model from pseudo-observations (Fig 5, Fig M in S1 File). This is not surprising since it uses extra variability information that is not available to the "trend" method. This extra information can provide a powerful constraint because models differ considerably in their representation of internal variability, based on sample estimates of the variability properties (Fig 4). The most striking improvement is obtained for the AMOC experiment while arguably the least improvement – for the AMOCIndex (Fig M in S1 File).

Another important metric is the factor $f$ that provides calibrated projections. This factor can be interpreted as a rough measure of model error relative to the next-closest inter-model differences in output space. The values feature a substantial range from 0.75 to 3.75 (Table 1). For experiments AMOC, AMOCIndex, and Korea_temp_long, $f$ is the same or similar for both methods. Thus, under our statistical model, the best dynamical models for both "trend" and "trend+var" experiments are approximately equally close to the "true" unobserved trends in the real system, in both calibration and projection periods. However, for the rest of the simulated data experiments the new method achieves a lower $f$. Here, the best model for the "trend+var" method is closer (more than twice as close for Korea_temp) to the "true" trend of the system, compared to the best model under the "trend" experiment, both in calibration *and* projection periods.

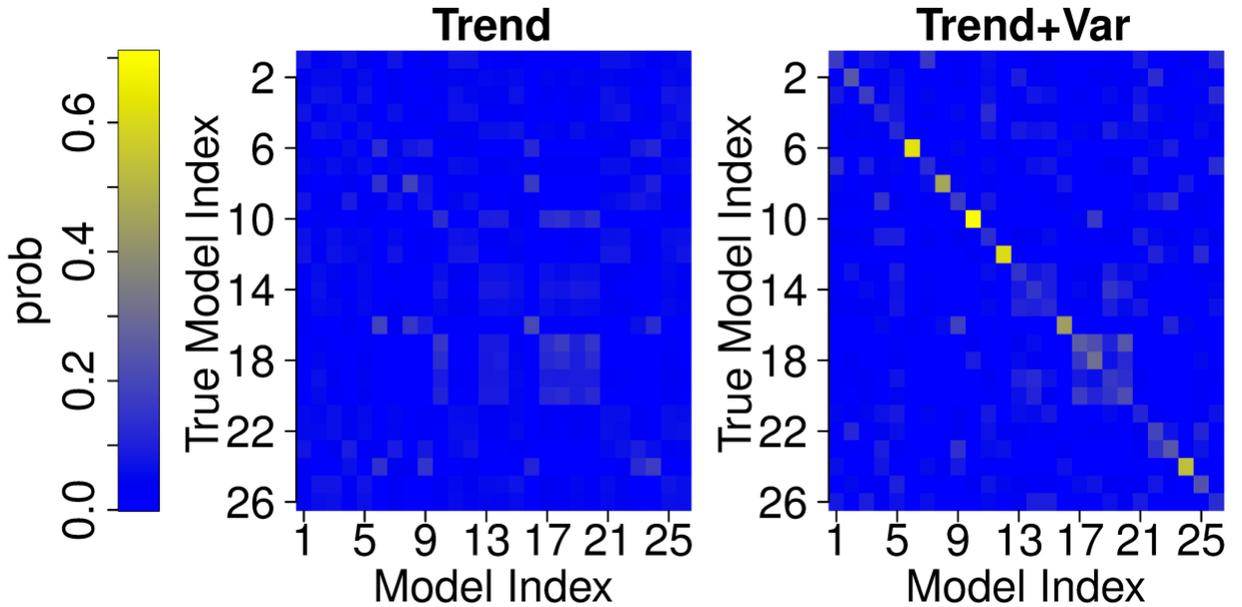

**Fig 5.** Model weights for Winter_SST cross-validation experiments. Rows represent different "true" models. Color represents model weight.

We now turn our attention to the question of future prediction. First, it is worth noting that we do not find a significant bias between projections and "true" model output in any of our leave-one-out cross-validation experiments. The new method tends to improve in terms of the mean 90% credible interval width as well as mean absolute bias of the mean (Table 1, Figs 6, 7, Figs N-Q in S1 File). Specifically, in the Korea_temp experiments, the forecast 90% credible intervals on average sharpen by about 25%. For some cases (e.g., models 3 and 22 of the Korea_temp experiment), the improvements are particularly dramatic, featuring a drastic sharpening of the pdf and a strong reduction in the 90% credible intervals, with a low bias; Figs P and Q in S1 File). The only cases with no improvement are AMOCIndex, and corresponding AMOCIndex_obs (Table 1). We note that these experiments rely on the same model output. They also use a weaker historical climate forcing during the projection period, whereas other experiments use stronger RCP8.5 future forcing. It is worthwhile noting that the experiments with the improvement boast a visual relationship between sample estimates of variability properties and future changes (Fig 4). Specifically, models with higher innovation standard deviation tend to produce higher summer mean maximum temperature warming in the Korean temperature experiments. A positive relationship between standard deviation and future temperature change has been previously found in previous work for many regions [6]. The relationships for the AMOC experiment are different: future AMOC slowdown appears to be stronger for models with higher autocorrelation and low normalized innovation standard deviation. In the Winter_SST experiment, the relationships also involve both variability properties: higher $\check{\sigma}_{M,i}$ and low $\check{\rho}_{M,i}$ in the models are associated with smallest future warming. Thus, we speculate that the degree of improvement may be related to the strength of statistical relationships between the variability parameters and future change. Testing this hypothesis is left to future work. There can be considerable shifts in the pdf between the "trend" and "trend+var" method (Figs N-Q in S1 File). This is consistent with the fact that additional fluctuation data can provide a relatively independent constraint on the model weights.

We note that the improvement in performance by the "trend+var" method is not caused by any increase in number of parameters resulting in overfitting. The overall statistical model for the projections is the same in both cases: a weighted mixture of pdfs from individual models. The increase in skill is due to better estimation of individual model weights $w_i$ in the "trend+var" model through using new variability data constraints on the models.

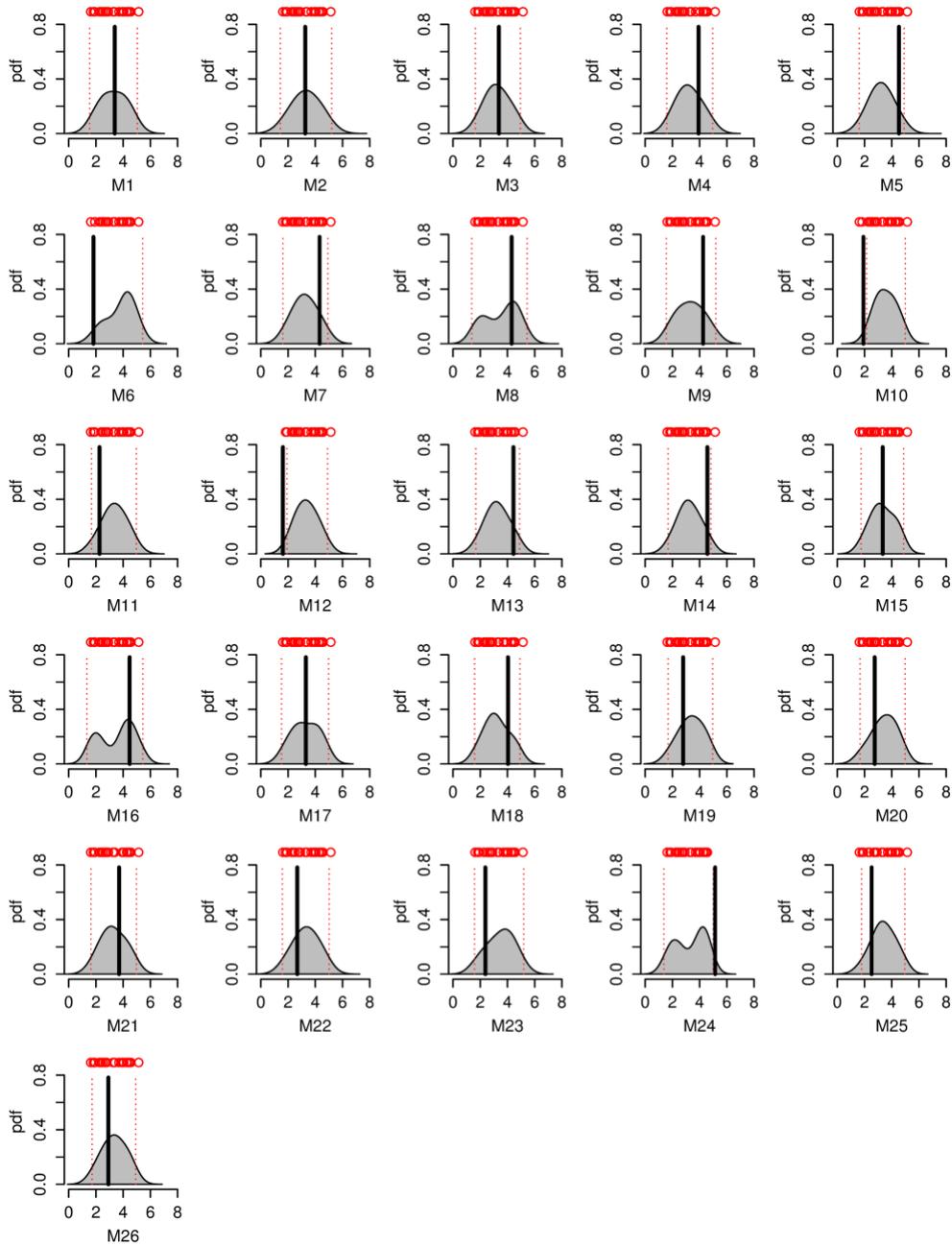

**Fig 6.** Probabilistic projections for winter East Sea surface temperature change from 1941-2000 to 2061-2100 [K] under the RCP8.5 emissions scenario for the "trend only" Winter_SST cross-validation experiment. Subplots differ in the assumed "true" model. Red circles are deterministic projections from each remaining model, red dotted lines are 90% posterior credible intervals. Black lines are changes from the "true" models.

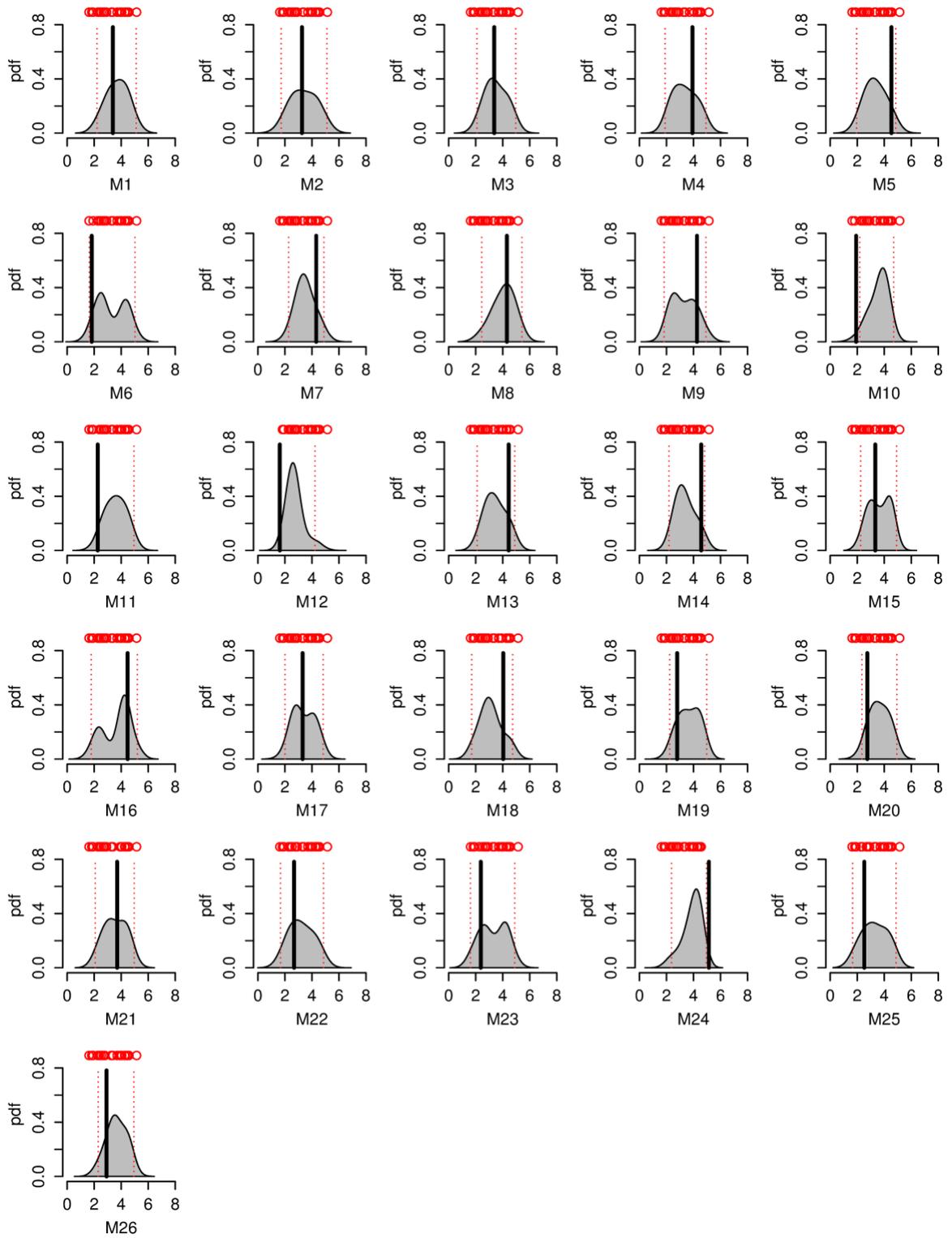

**Fig 7.** As Fig 6, but for "trend+var" experiment implementing the new proposed method

# 5 Real-Case Application: Projecting Korean Summer Mean Maximum Temperature

We now apply both the "trend" and "trend+var" methods to make projections of Korean summer mean maximum temperature. Specifically, we use 29 GCMs from Coupled Model Intercomparison Project phase 5 (CMIP5, [40]) model runs (the same model set as for the Korea_temp experiment). The models are weighted using 1973-2005 station observational data provided by Korean Meteorological Administration (KMA) weather stations [38,53]. We apply simple area average to daily mean maximum temperatures from the stations before calculating summer mean values. We use this short period because it has the best observational coverage, however to provide a liberal estimate of the uncertainty we take model error expansion factors *f* from the corresponding longer-period Korea_temp_long experiments. Future changes (2081-2100 minus 1973-2005) under the RCP8.5 emissions scenario [42] are presented in Fig 8. The results show (a) notably higher projected warming and (b) considerable reduction of the low-warming (< 2 K) tails after the variability weighting. Specifically, the mean increases from 4.9 K to 5.6 K, and the 5$^{th}$ percentile from 1.8 K to 3.2 K. The new projection mode leaps from 5.3 K to 6.6 K (Table 2). In addition, the 90% credible interval shrinks from 5.5 K to 4.3 K (22 % reduction).

**Table 2.** Summary of Korean summer mean maximum temperature change probabilistic projections from 1973-2005 to 2081-2100 under the RCP8.5 emissions scenario from the "trend" and "trend+var" methods.

| Experiment | Mean | Median | Mode | 90% Credible Interval |
|---|---|---|---|---|
| | | | | |
| Trend | 4.9 K | 5.0 K | 5.3 K | (1.8, 7.3) |
| Trend+var | 5.6 K | 5.9 K | 6.6 K | (3.2, 7.5) |

# 6 Discussion

Here we present a novel method "trend+var" to weight models of complex dynamical systems by their skill at representing both autocorrelated variability and trend in observations. The key step is association of two statistical models with each dynamical model: a trend statistical model, and a variability statistical model. The component submodels are weighted separately using relevant observations, and then the weights are combined. The combined weights are used to make weighted probabilistic multi-model projections. In a series of cross-validation experiments, we show that the new method appears to better identify the "true" model compared to the trend-only weighting method ("trend"). The new method also tends to perform better in terms of mean 90% posterior credible interval and mean absolute bias. Our analysis deviates in some aspects from the traditional Bayesian framework, in order to avoid making difficult-to-justify parametric assumptions about model error, and to alleviate potential overconfidence in one-at-a-time cross-validation experiments.

Applying the new method to the real case of projecting Korean summer mean maximum temperature change by the end of this century considerably increases future projections. These projections are more informative than from the "trend" method because they use the additional variability and short-term memory (quantified by the lag-1 autocorrelation coefficient) information from both models and observations. Since the BMA predictive model is the same (Eq. 14), the increase in skill is not due to an increased number of parameters, but is derived purely through better estimation of model weights. Recent work has found correlations with absolute values of up to approximately 0.8 between present-day interannual summer temperature sample standard deviation in global and regional climate models, and long-term future mean and/or variability changes for some regions [6,25]. This suggests that historical variability in those regions may provide a valuable constraint on the models. Applying the method to those regions should be considered for future work.

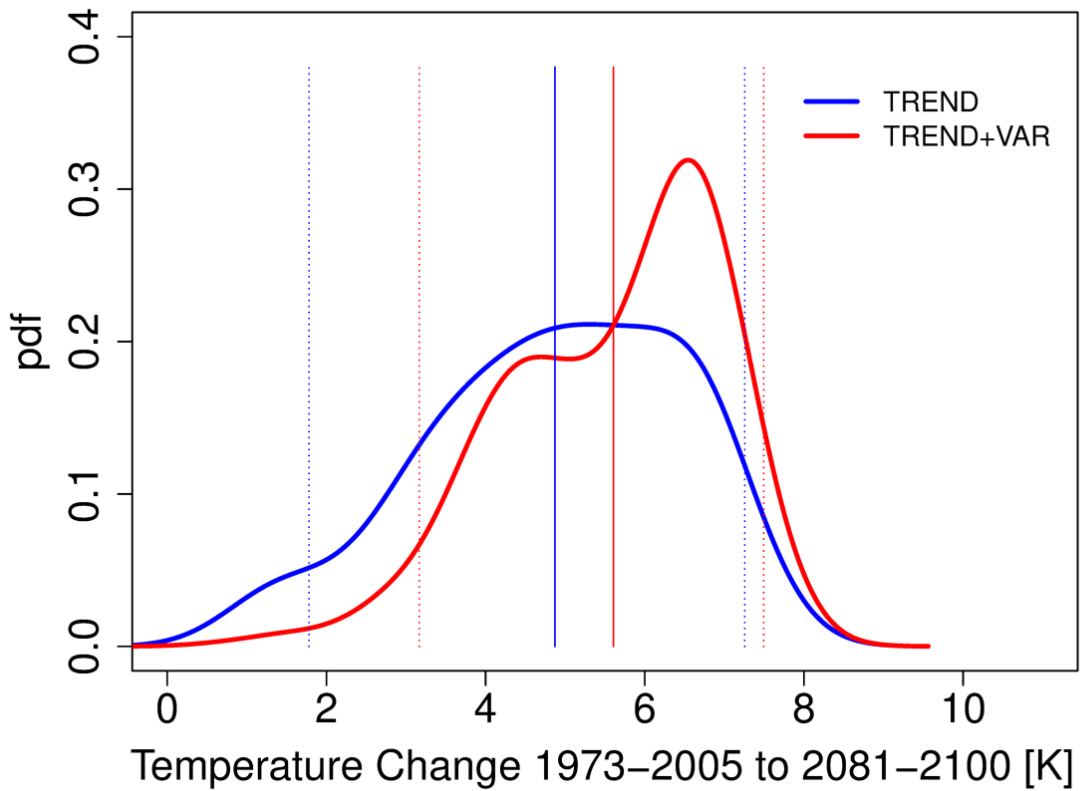

**Fig 8.** Probabilistic projections of summer mean maximum temperature change 1973-2005 to 2081-2100 over Korea under the RCP8.5 emissions scenario using "trend" and "trend+var" methods. Vertical lines are the means and the 90% posterior credible intervals.

It is worth discussing differences between this study and previous Bayesian work. Here we for the first time implement a quasi-Bayesian statistical method that weights models by their performance in terms of both trend, variability, and short-term memory (as quantified by the lag-1 autocorrelation) for a relatively general case: arbitrary (potentially non-linear) trend function and red noise variability. The method can be extended to more complex variability structures. Model weights are obtained by constraining the method with calibration period observations, while a parameter controlling model error assumptions is calibrated using cross-validation experiments. Some prior work does also incorporate variability into model weights [35], however their method has so far been demonstrated on a simple case: serially uncorrelated variability, and a linear mean function. Other studies [3,4,6] also incorporates variability into the analyses. However, these studies do not actually use variability performance to weight the models and ignore autocorrelation skill. Unlike previous work, we do consider autocorrelation, which is a common feature of variability in many observed and modeled processes [54–56].

# 7 Caveats

Our study is subject to several caveats. First, the anomalies around the long-term trend, as well as model-observational residuals are assumed to be red noise processes. However, our framework can be extended to more general cases in the future. We compare the spectra of model anomalies (normalized in the AMOC experiment) for each model and experiment to the 90% confidence intervals for the corresponding AR1 process spectra, based on 1000 random realizations (Figs R-T in S1 File). Relevant comparison for the AMOCIndex experiment is shown in Fig 4 of a preceding study [9]. These results indicate that AR1 process is a reasonable approximation to the internal variability for these systems. Second, when combining the weights of the variability and trend submodels we are assuming independence. While this assumption appears to be generally reasonable here, it may not apply for other datasets. Incorporating dependence should be considered in future studies. Thus, our method is expected to be ideal for cases where there is at least some relationship between present-day variability and future changes, yet the relationship between present-day trends and variability in the models is sufficiently weak to justify the independence assumption we make here. Third, by using a common error expansion factor $f$ for the internal variability, trend submodel errors, as well as for the forecasts, we are assuming the magnitudes of errors in these three components are linked. A way forward in subsequent work may be to assume different $f$ for trend and variability. The best $f$ values could then be found using constrained optimization (optimizing future performance metrics while constraining coverage to be correct). This is beyond the scope of this study. Fourth, when sampling future internal variability, we do not consider the uncertainty in the AR1 parameters of the anomalies. However, as explained in Section 3, we calibrate our method to account for potential overconfidence by scaling the magnitude of the model errors. Other caveats include the simplicity of the future model bias and of the cross-validation experiments, as well as no explicit representation of observational error. For the future Korean temperature projections, the high density of observational network mitigates some of these concerns, as random errors are expected to decrease after averaging across many stations. In addition, theoretically if modelled and observed data from multiple regions are used together in a cross-validation framework, the observational error will be implicitly incorporated into the analysis after nudging the $f$ parameter. Nonetheless, an explicit representation of observed error should be considered in the future.

While the focus on this paper is on the statistical weighting methodology by trend and variability performance, the simplicity of the decomposition into trend and variability (e.g., lowess method or linear detrending) deserves mention. The nonlinear trends discussed here may include residual contributions from long-term internal climate variability. However, this can be handled by the trend-weighting part of the method since this part accounts for long-term model error [35]. The unfiltered long-term variability in each model can be simply considered as part of this long-term model error. Previous work provides examples of using a more sophisticated decomposition [18]. Improving the decomposition methodology is beyond the scope of the paper, and is subject of future work.

This work assumes stationarity of model weights: if a model is correct during the calibration period, it is also assumed to be correct in the validation period. This is a standard assumption of the BMA method [1,5,20,21,35].

Notably, this work does not properly confront the issue of model dependence (e.g., the fact that models coming from the same research group, or models with similar outputs are dependent in the general sense of the term) [12,57–60]. This needs to be addressed in future work.

The best new datasets to apply the method to are the ones either with many regions, or with repeated experiments, and where a long calibration period can be split into two subperiods. In this case method performance can be systematically assessed using real observations in cross-validation experiments, and $f$ can be properly calibrated. However, any assumption about $f$ under new conditions is inherently untestable. Hence, we recommend including equal weights projections along with projections from this (or any other) weighting scheme. In the absence of many regions, and with only short time series available, one has to resort to simulated cross-validation experiments using calibration, projection, and projection reference period model output to calibrate the method. In such cases, if models share common errors, the real value of *f* may be higher than estimated.

# 8 Conclusions

We present a statistically-rigorous novel method to weight multiple models of stochastic dynamical systems by their skill at representing both internal variability (including autocorrelation) and a nonlinear trend of a time series process, and to make predictions of system change under new conditions. The weight is interpreted as a likelihood of a dynamical model being adequate at capturing both trend and variability aspects of the process. This is a particularly important diagnostic given the broad relevance of variability (e.g., variability can affect extreme events such as heat waves and droughts in climate science). We show that the proposed method tends to better identify "true" models in a suite of leave-one-out cross-validation experiments compared to a typically-used trend-only BMA weighting method. The new method also tends to improve forecasts, as judged by the mean 90% credible interval width and mean absolute bias. This has important implications specifically for multi-model climate projections. Applying the method to project Korean summer mean maximum temperature changes over this century considerably increases future projections. Specifically, the mode of 1973-2005 to 2081-2100 warming under the RCP8.5 emissions scenario increases by 1.3 K to 6.6 K, while the mean shifts from 4.9 K to 5.6 K. Furthermore, the pdf becomes 22% sharper as measured by the 90% posterior credible interval.

# Acknowledgments

For their roles in producing, coordinating, and making available the CMIP5 model output, we acknowledge the climate modeling groups, the World Climate Research Programme's (WCRP) Working Group on Coupled Modelling (WGCM), and the Global Organization for Earth System Science Portals (GO-ESSP). A portion of model outputs used here has been obtained from the German Climate Computing Centre (DKRZ), funded through the Federal Ministry for Education and Research. Jong-Soo Shin provided technical assistance with extracting Korean and East Sea temperature model output. We are thankful to Cameron Farnsworth for providing comments on the manuscript, and to Soong-Ki Kim for thought-provoking discussions.

# Supporting Information for "Accounting for skill in trend, variability, and autocorrelation facilitates better multi-model projections: application to the AMOC and temperature time series"

Projections accounting for model trend and variability skill

**Roman Olson[1,2,3], Soon-Il An[1], Yanan Fan[4] and Jason P. Evans[5]**

[1]Department of Atmospheric Sciences, Yonsei University, Seoul, South Korea

[2]Center for Climate Physics, Institute for Basic Science, Busan, South Korea

[3]Pusan National University, Busan, South Korea

[4]School of Mathematics and Statistics, UNSW Australia, Sydney, NSW, 2052, Australia

[5]Climate Change Research Centre and ARC Centre for Excellence in Climate Extremes, UNSW Australia, Sydney, NSW, Australia

**Contents of this file**

>Texts A and B
>Figs A to T
>Tables A to B

**Introduction**

Supporting Texts contains information on details of the statistical methodology. Text A provides technical details on weighting trend models, while Text B discusses the implementation of the variability model weighting.

**Text A. Technical details on weighting the trend models**

For the Monte Carlo integration to get trend submodel weights, we use 100,000 samples for all experiments except "trend" Korea_temp, AMOCIndex and AMOCIndex_obs which use 1,000,000 samples. The real-case Korean temperature projections also use 1,000,000 samples. When we repeat Korea_temp experiments with a different number of samples (1,000,000 for "trend+var" and 100,000 for "trend"), the performance metrics for these experiments are virtually identical. This suggests that 100,000 samples are enough to reasonably estimate method performance. We employ uniform priors for $\sigma$ on [0, 5], and for $\rho$ on [-1, 1] for the AR1 properties of the internal variability during the trend weighting in all experiments.

**Text B. Implementation of the variability model weighting**

For the Monte Carlo integration to get variability submodel weights, we use 10,000 samples for all relevant experiments. The real-case Korean temperature projections use 100,000 samples. Since the results for the performance metrics are virtually identical for the longer Korea_temp "trend+var" experiment described in Text A (which uses 100,000 samples for variability weights estimation), we deduce that 10,000 samples is a reasonable number. When sampling $\boldsymbol{\theta}_y^V$ we set autocorrelations with absolute values of 0.999 or higher to ±0.999 for numerical stability reasons. Likewise, we restrict ourselves to positive innovation standard deviations by setting all values below $0.01 \times \min(\boldsymbol{\breve{\sigma}}_M)$ to $0.01 \times \min(\boldsymbol{\breve{\sigma}}_M)$. Here $\boldsymbol{\breve{\sigma}}_M$ is a vector of all standard deviation summary statistics from all dynamical variability models: $\boldsymbol{\breve{\sigma}}_M = (\breve{\sigma}_{M,1}, \ldots, \breve{\sigma}_{M,k})$.

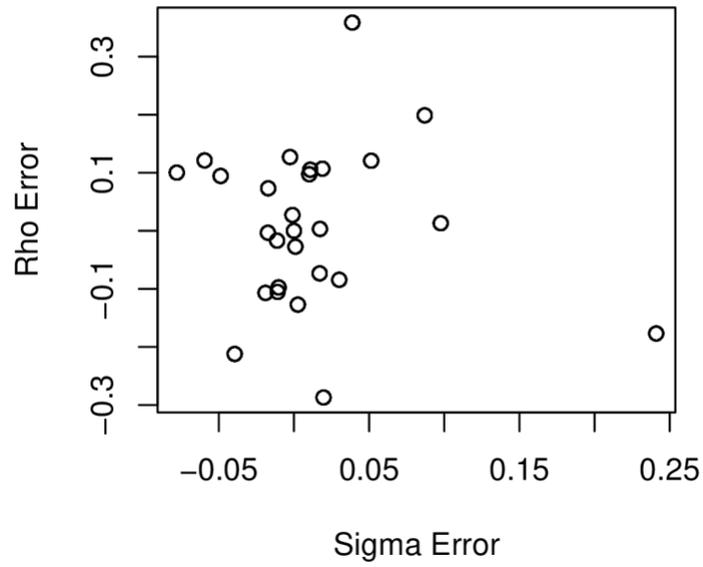

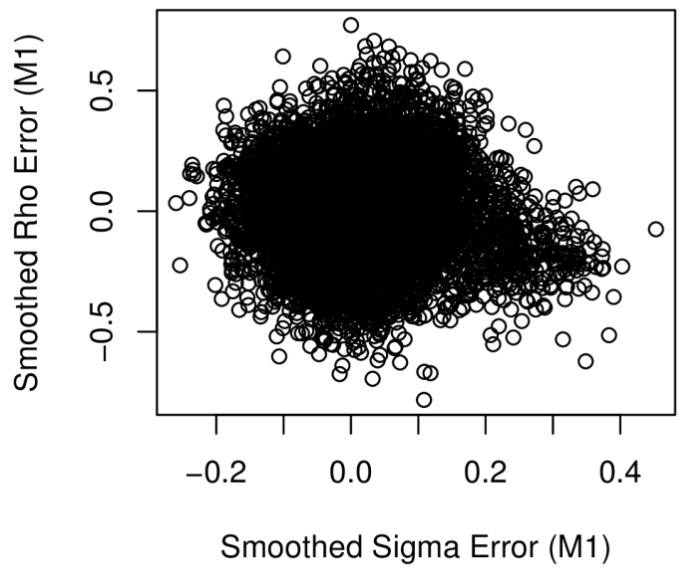

**Fig A.** (top) 26 samples of $f\varepsilon^V = f(\varepsilon_\sigma, \varepsilon_\rho)$ provided by next-closest model differences for the Winter_SST experiment, together with the (0,0) sample, (bottom) samples augmented by samples from the bivariate normal distribution with standard deviations set to 1/5 of the initial sample ranges, used in weighting the 1st variability model.

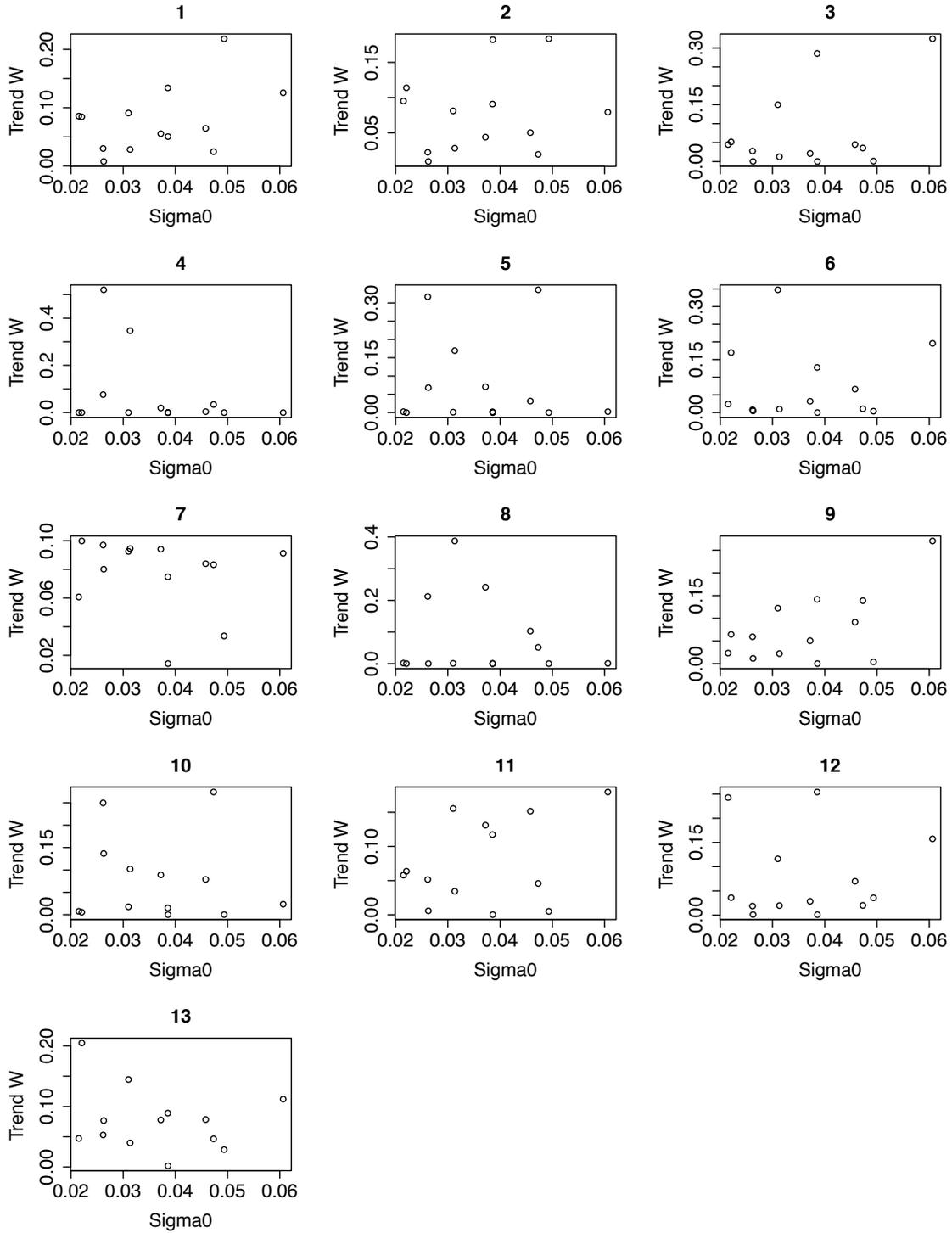

**Fig B.** Correlation between normalized innovation standard deviation summary statistics and trend weights $p(M_{T,i})$ for the AMOC experiment. Each panel corresponds to a different "true" model used as pseudo-observations.

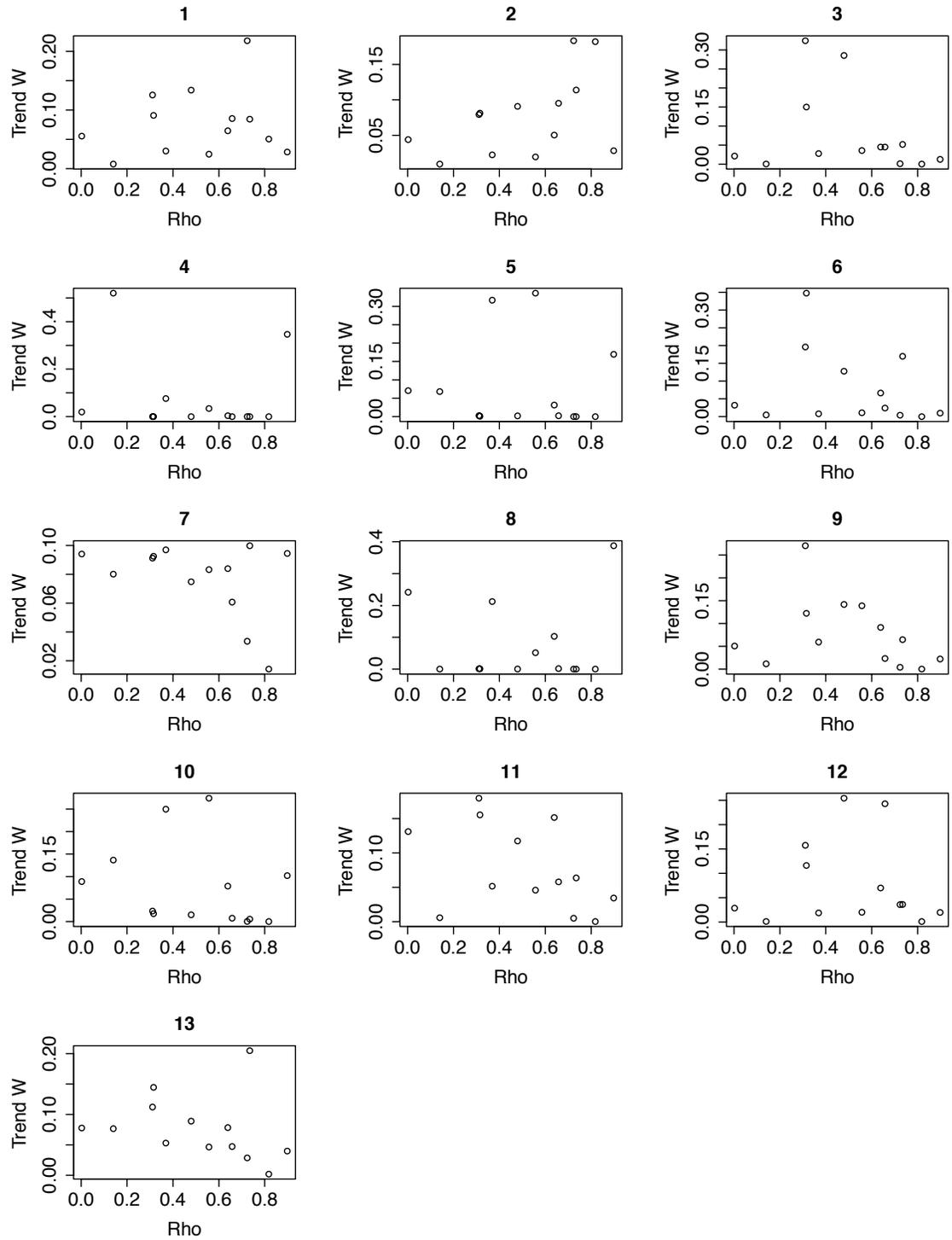

**Fig C.** Correlation between autocorrelation summary statistics and trend weights $p(M_{T,i})$ for the AMOC experiment. Each panel corresponds to a different "true" model used as pseudo-observations.

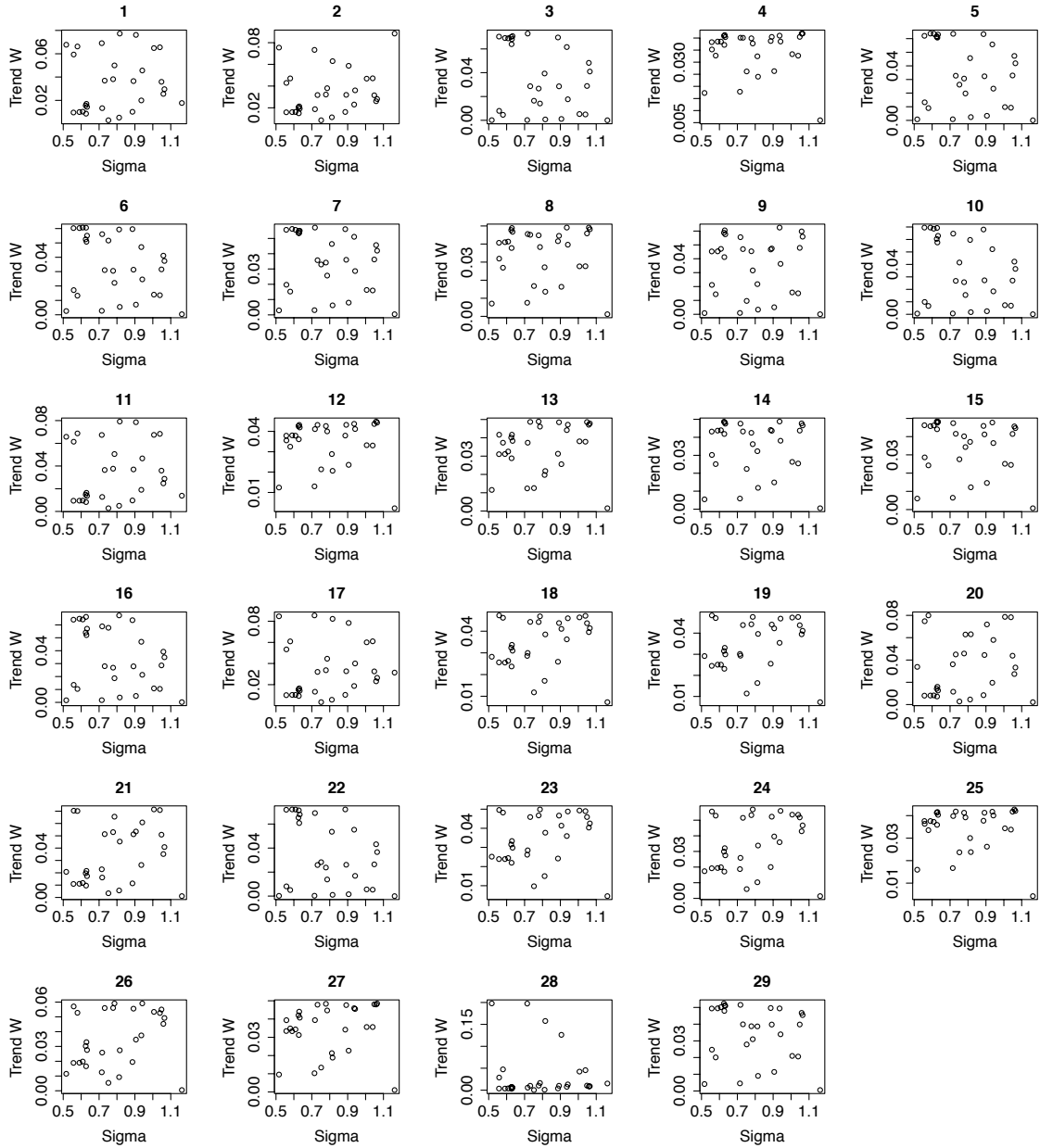

**Fig D.** Same as Fig B, but for un-normalized standard deviation [K] and the Korea_temp experiment.

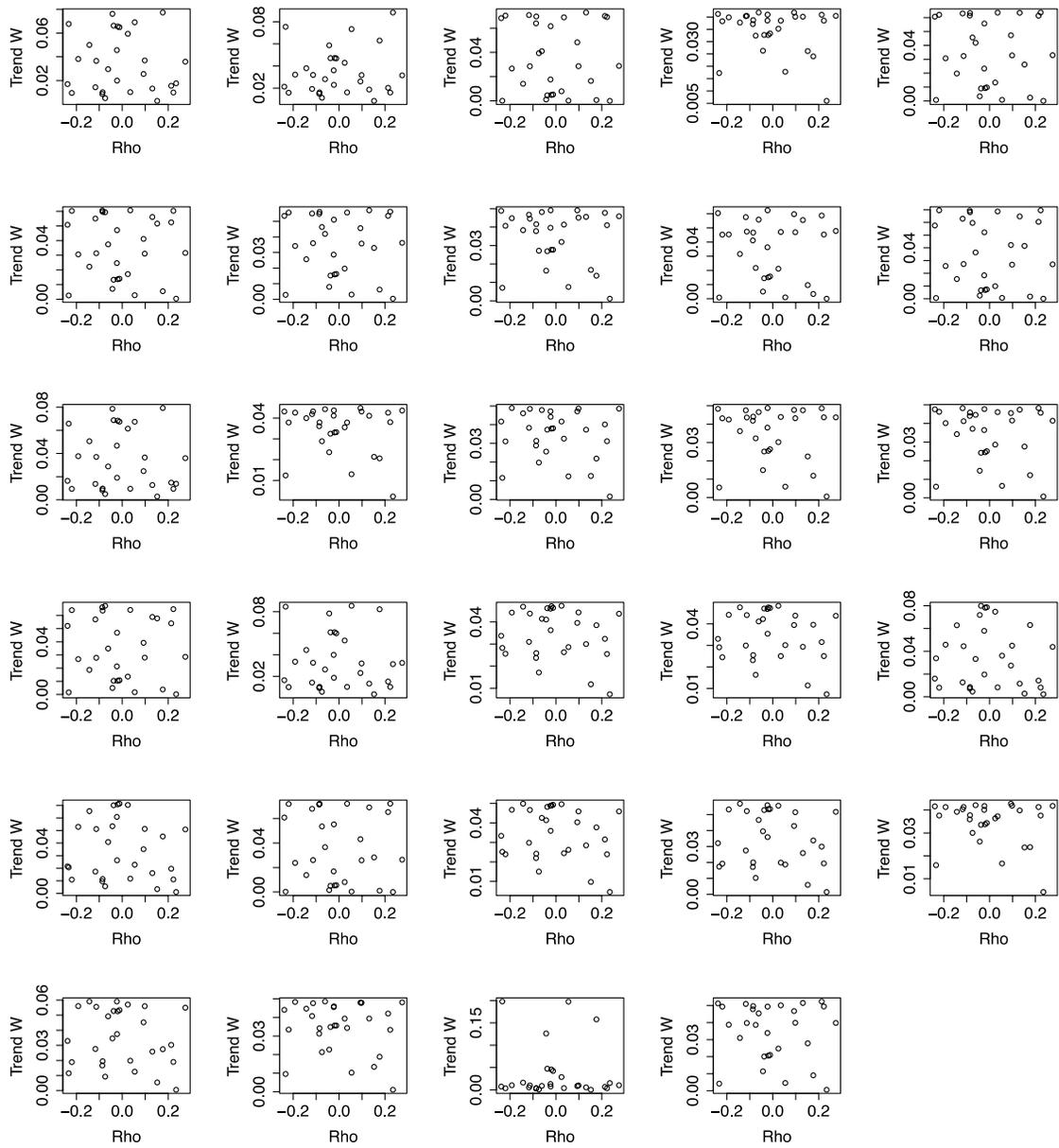

**Fig E.** Same as Fig C, but for the Korea_temp experiment.

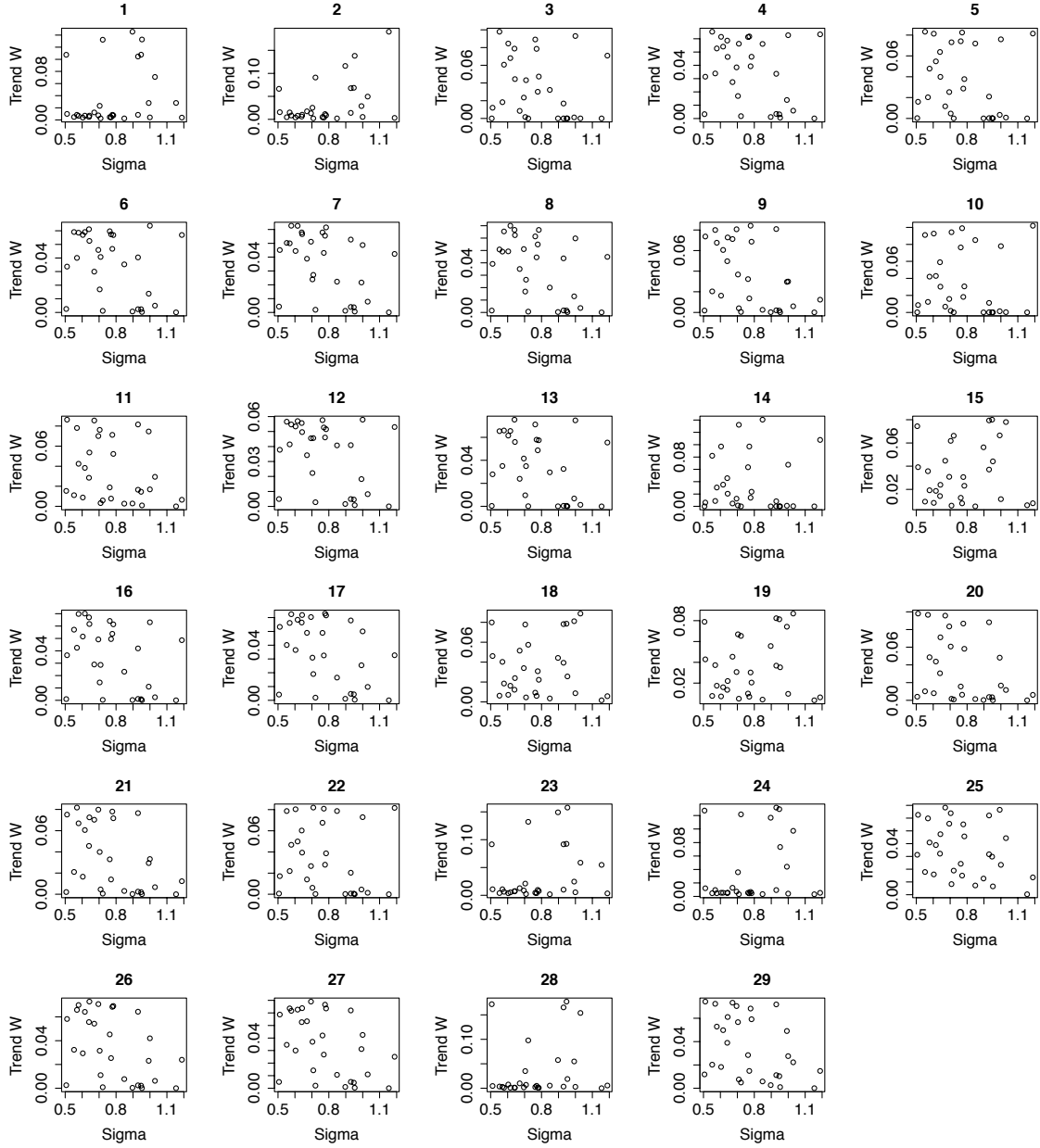

**Fig F.** Same as Fig B, but for un-normalized standard deviation [K] and the Korea_temp_long experiment.

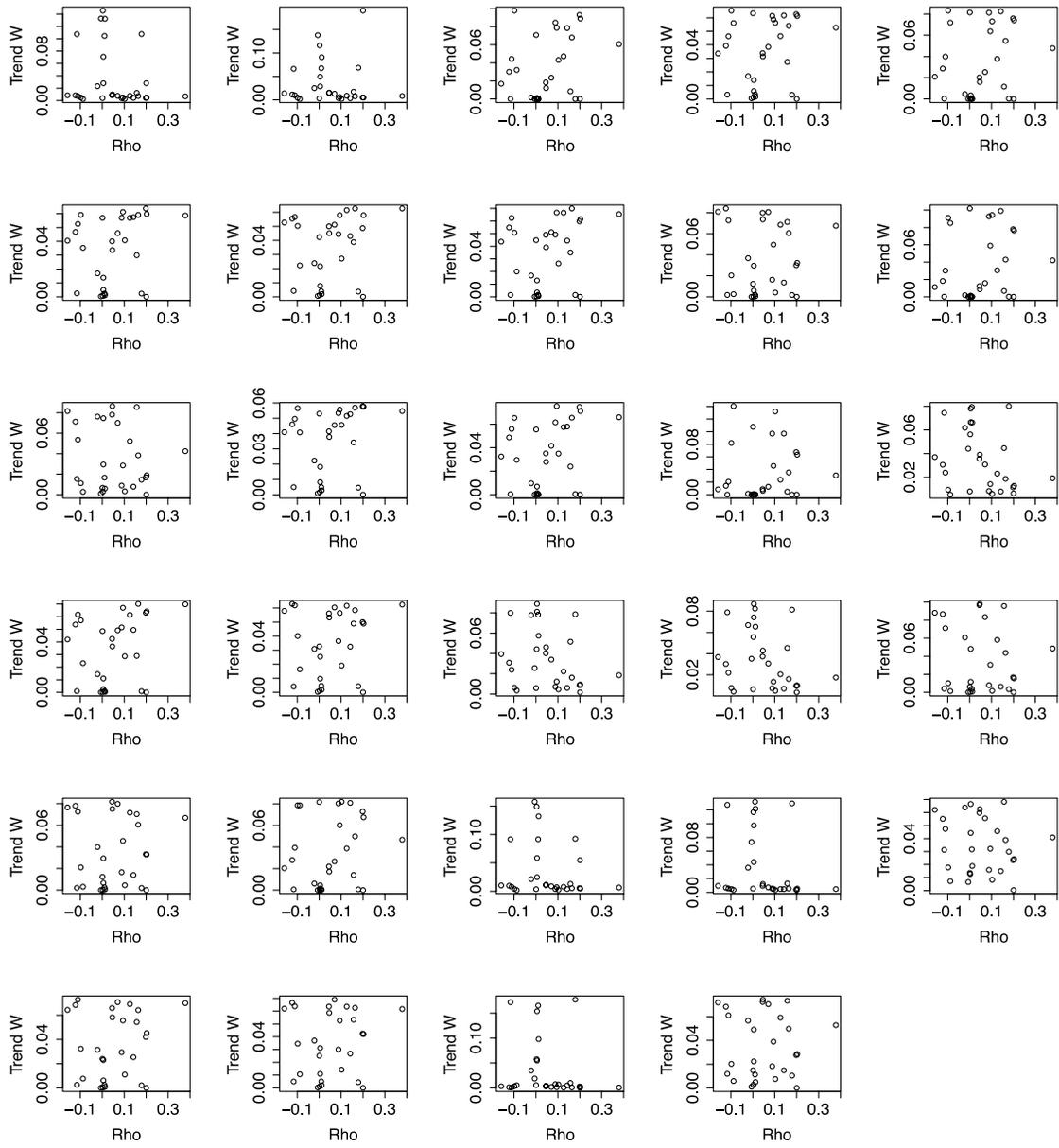

**Fig G.** Same as Fig C, but for the Korea_temp_long experiment.

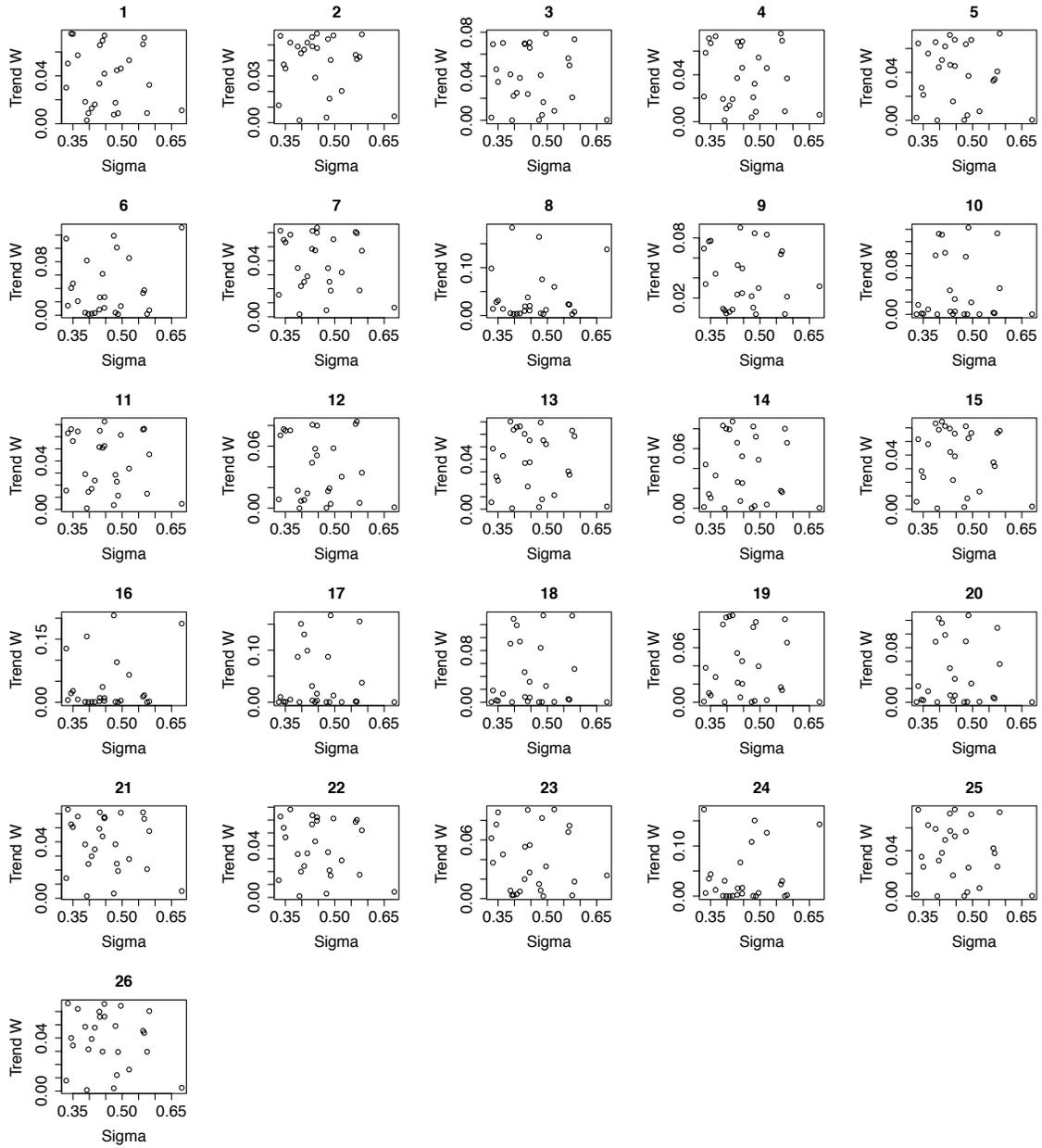

**Fig H.** Same as Fig B, but for un-normalized standard deviation [K] and the Winter_SST experiment.

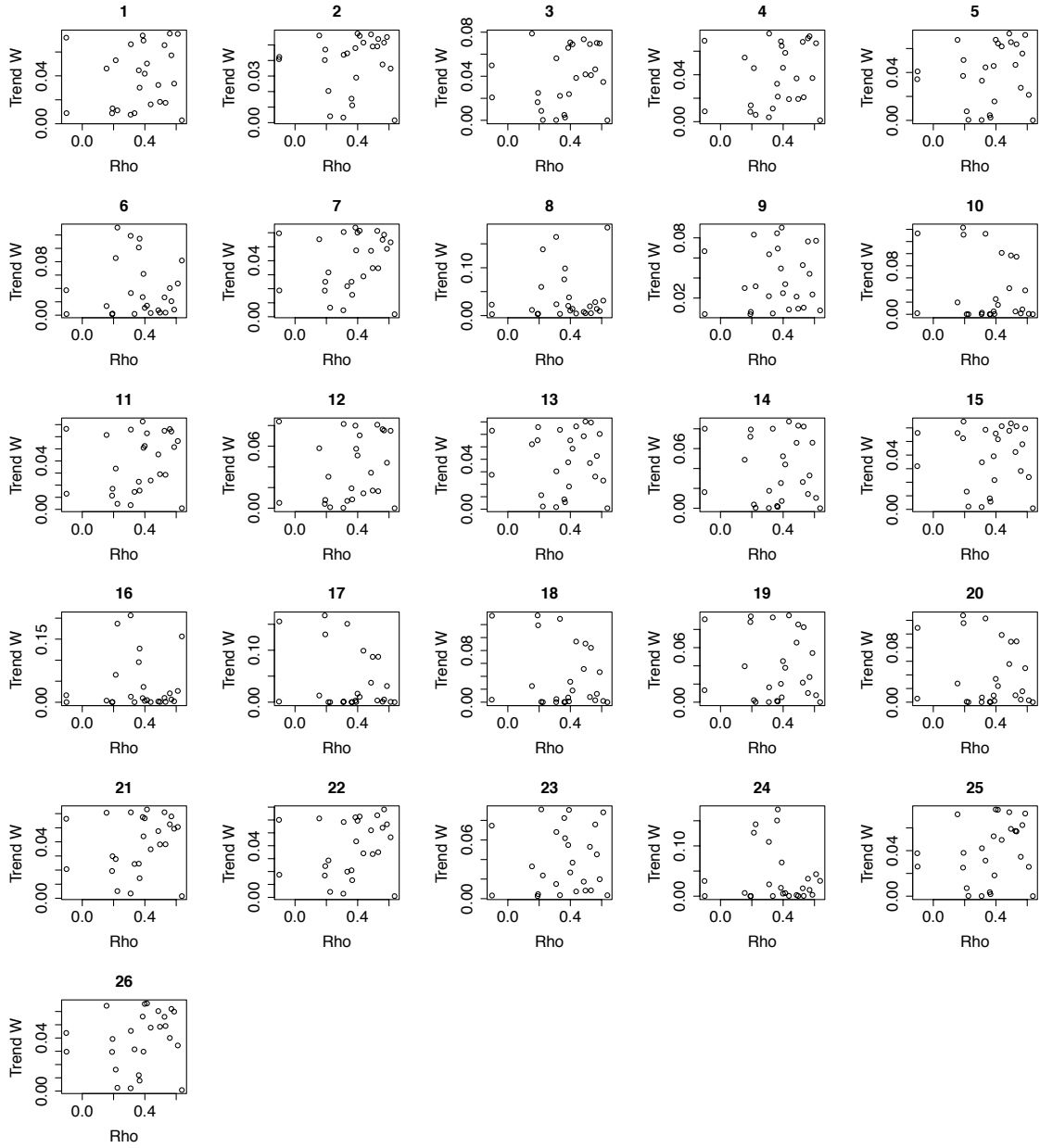

**Fig I.** Same as Fig C, but for the Winter_SST experiment.

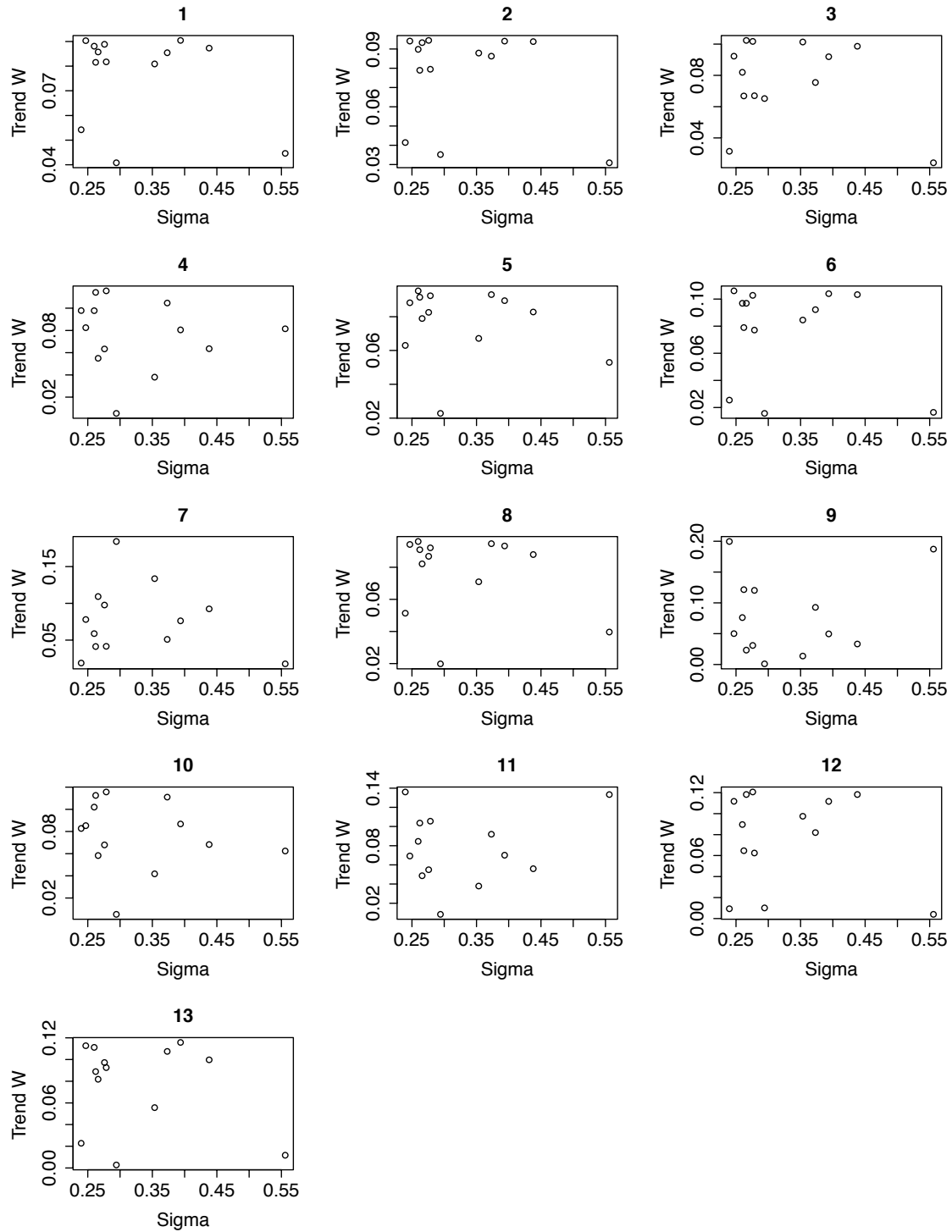

**Fig J.** Same as Fig B, but for un-normalized standard deviation [K] and the AMOCIndex experiment.

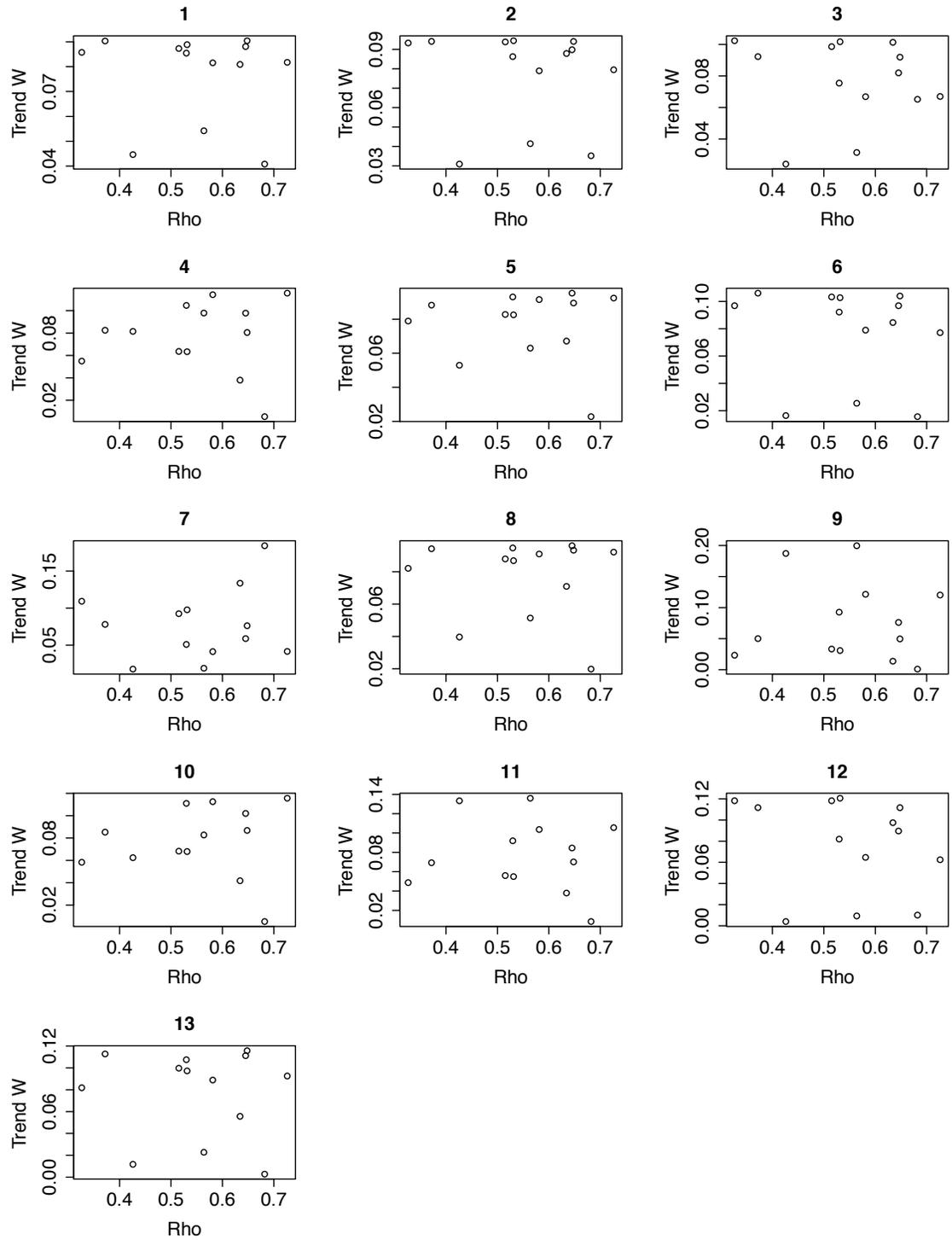

**Fig K.** Same as Fig C, but for the AMOCIndex experiment.

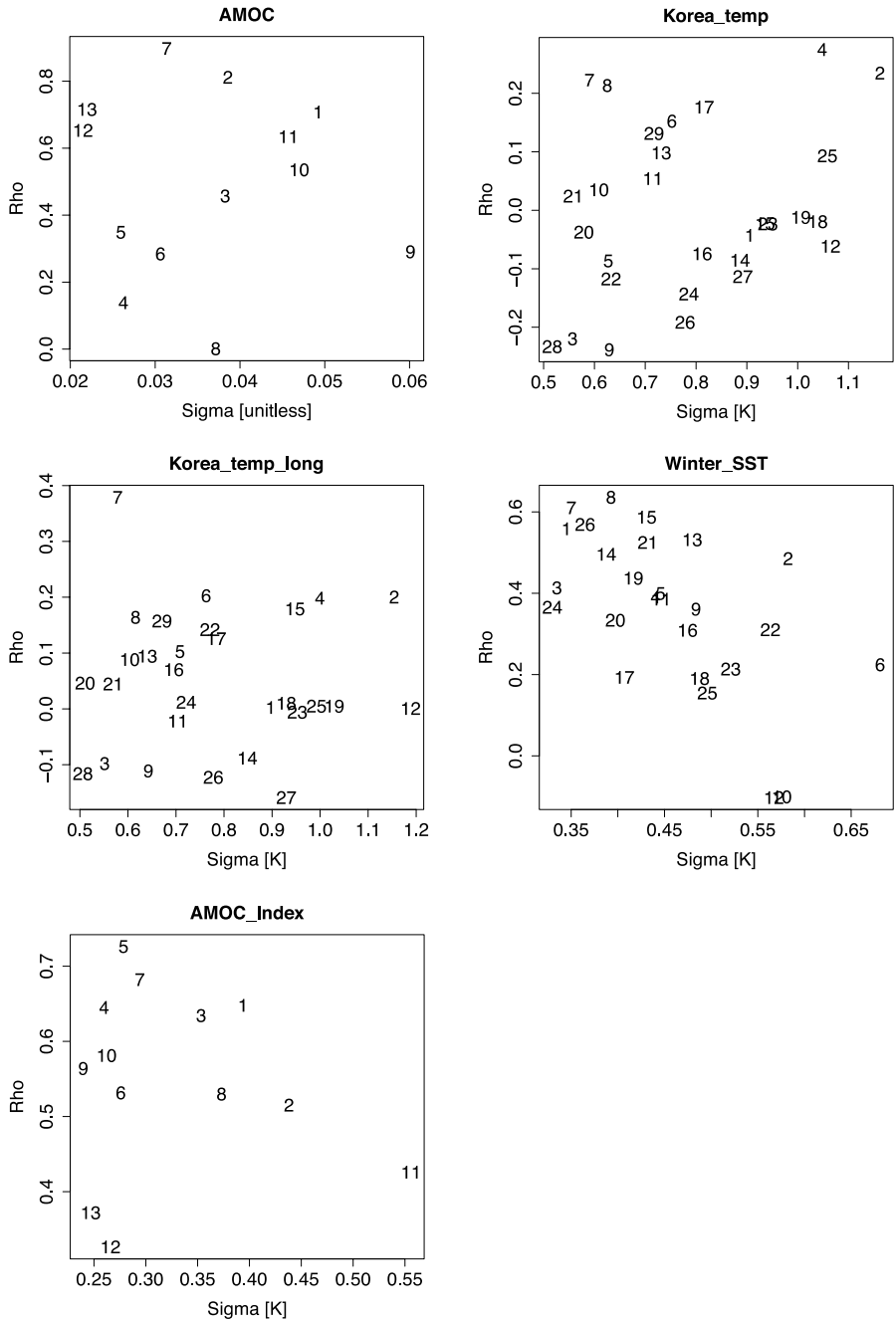

**Fig L:** Similar to Fig. 4, but with model numbers shown instead of future climate changes. For model numbers in AMOC and AMOC_Index experiments see previous work [1]. For the Korea_temp and Korea_temp_long experiments see Table A. For Winter_SST experiment see Table B.

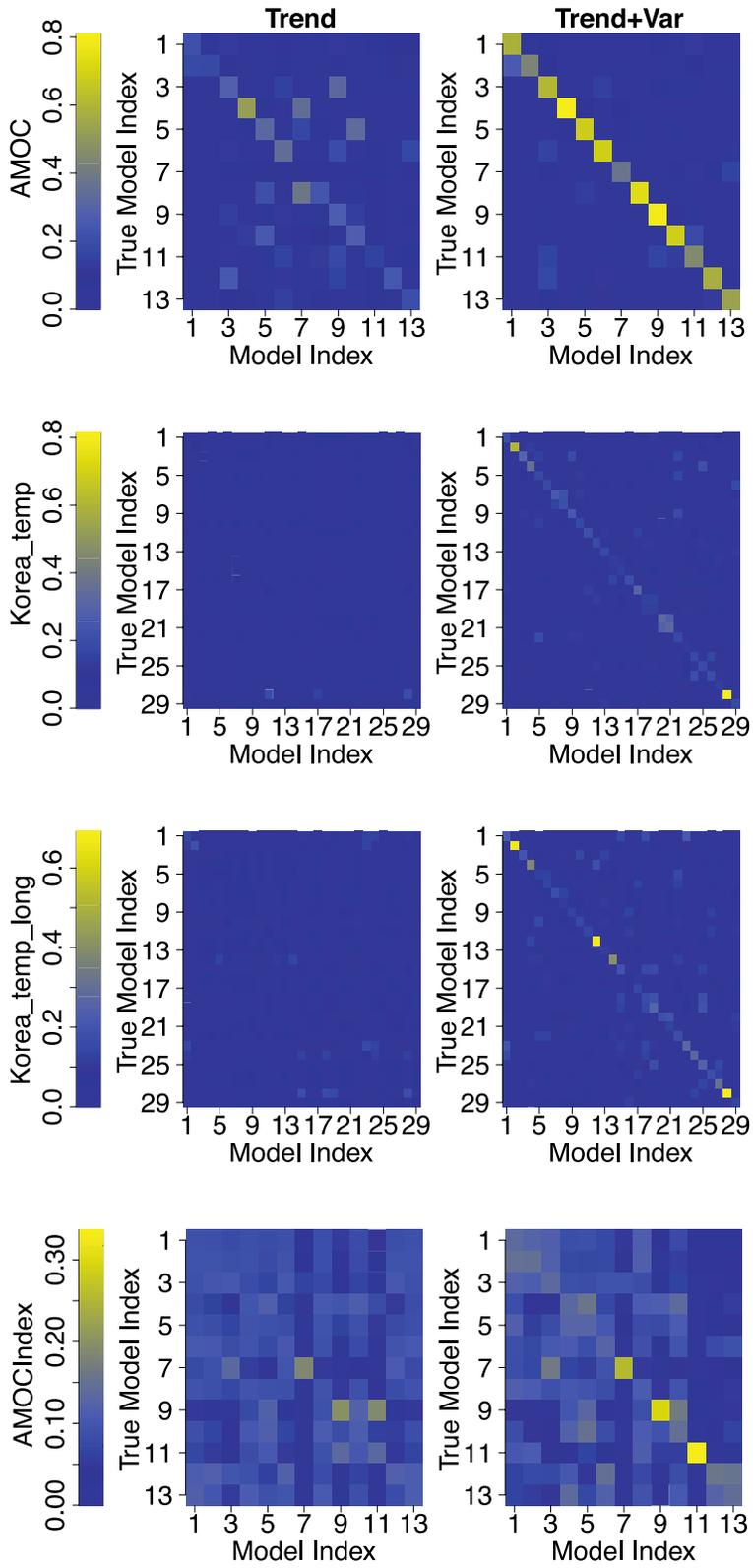

**Fig M.** Similar to Fig 5, but for the rest of one-at-a-time cross-validation experiments.

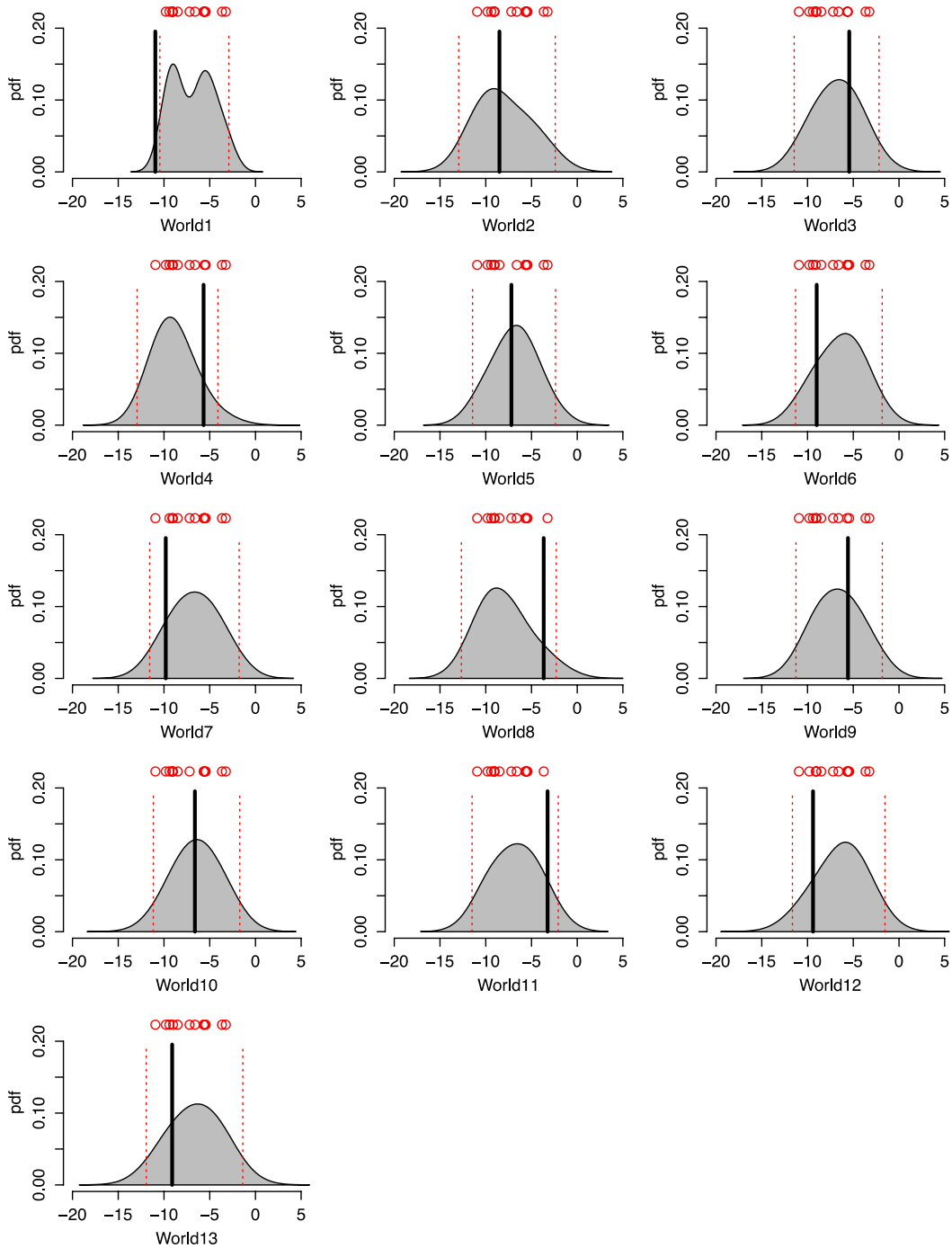

**Fig N.** Probabilistic projections for AMOC change from 1960-1999 to 2060-2099 [Sv] under the RCP8.5 emissions scenario for the "trend" AMOC cross-validation experiment. Subplots differ in the assumed "true" model. Red circles are deterministic projections from each model, red dotted lines are 90% posterior credible intervals. Black lines are changes from the "true" models.

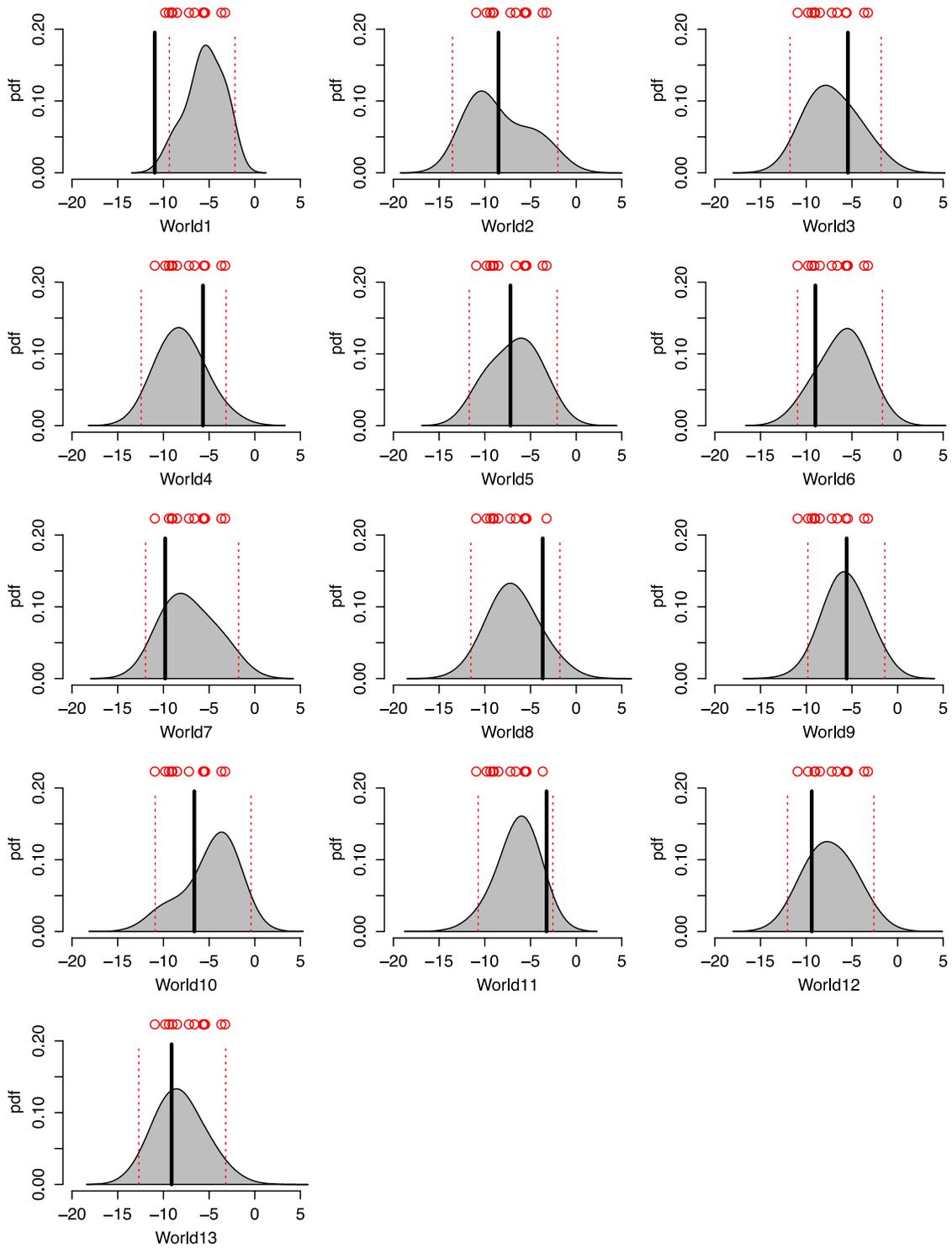

**Fig O.** Similar to Fig M, but for the AMOC "trend+var" experiment.

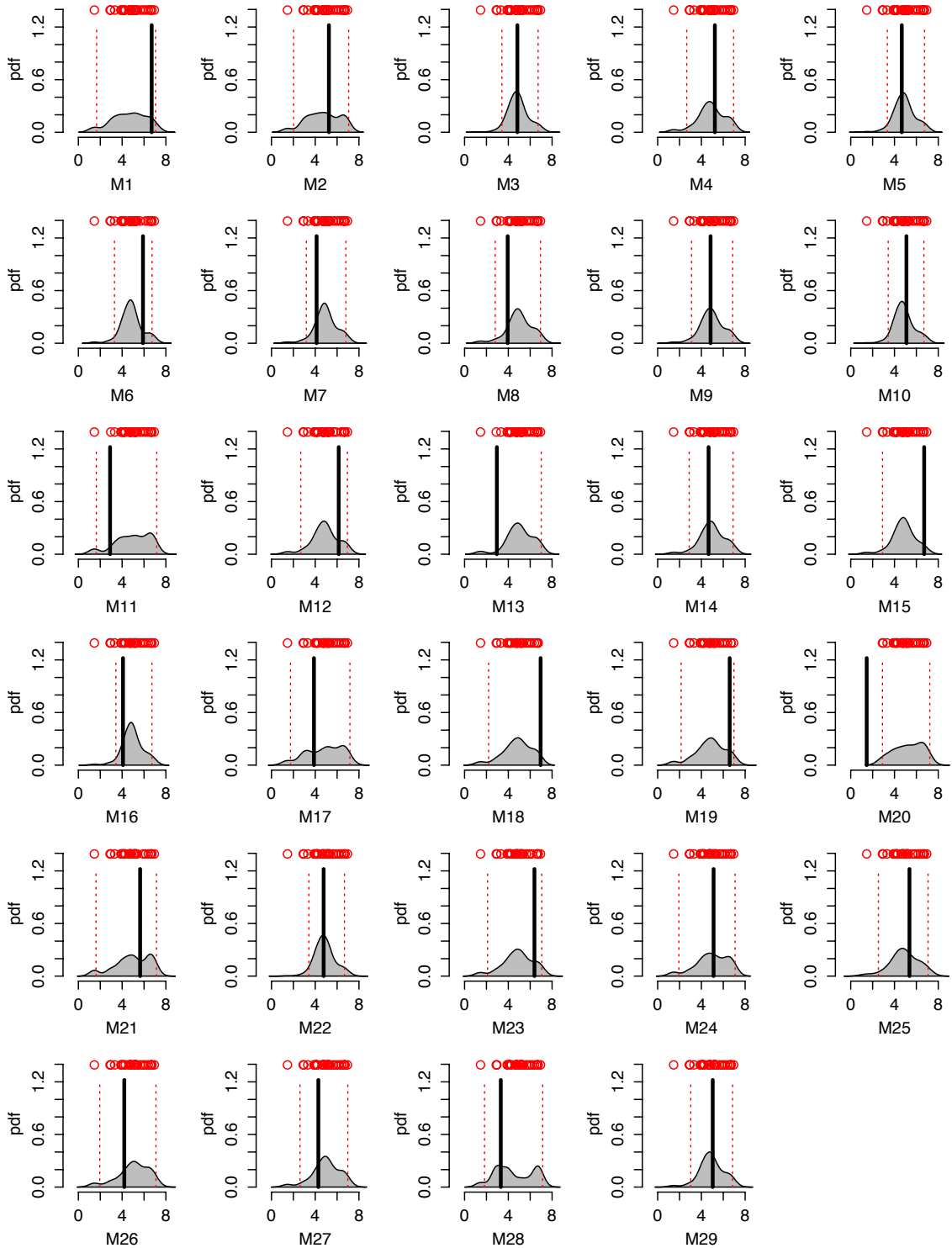

**Fig P.** Similar to Fig M, but for the Korea_temp "trend" experiment. The pdfs represent Korean JJA mean maximum temperature change from 1973-2005 to 2081-2100 [K] under the RCP8.5 emissions scenario.

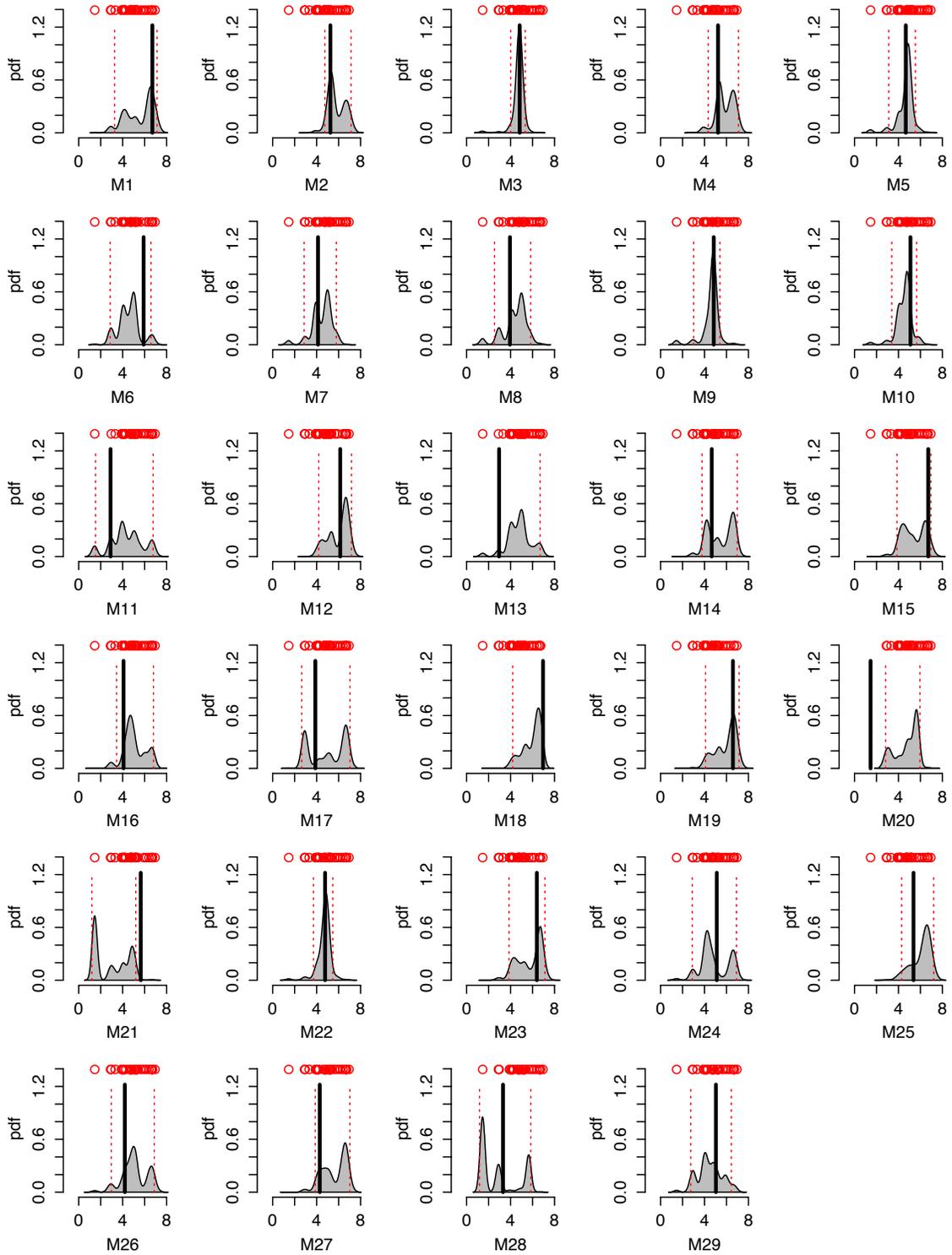

**Fig Q.** Similar to Fig O, but for the Korea_temp "trend+var" experiment.

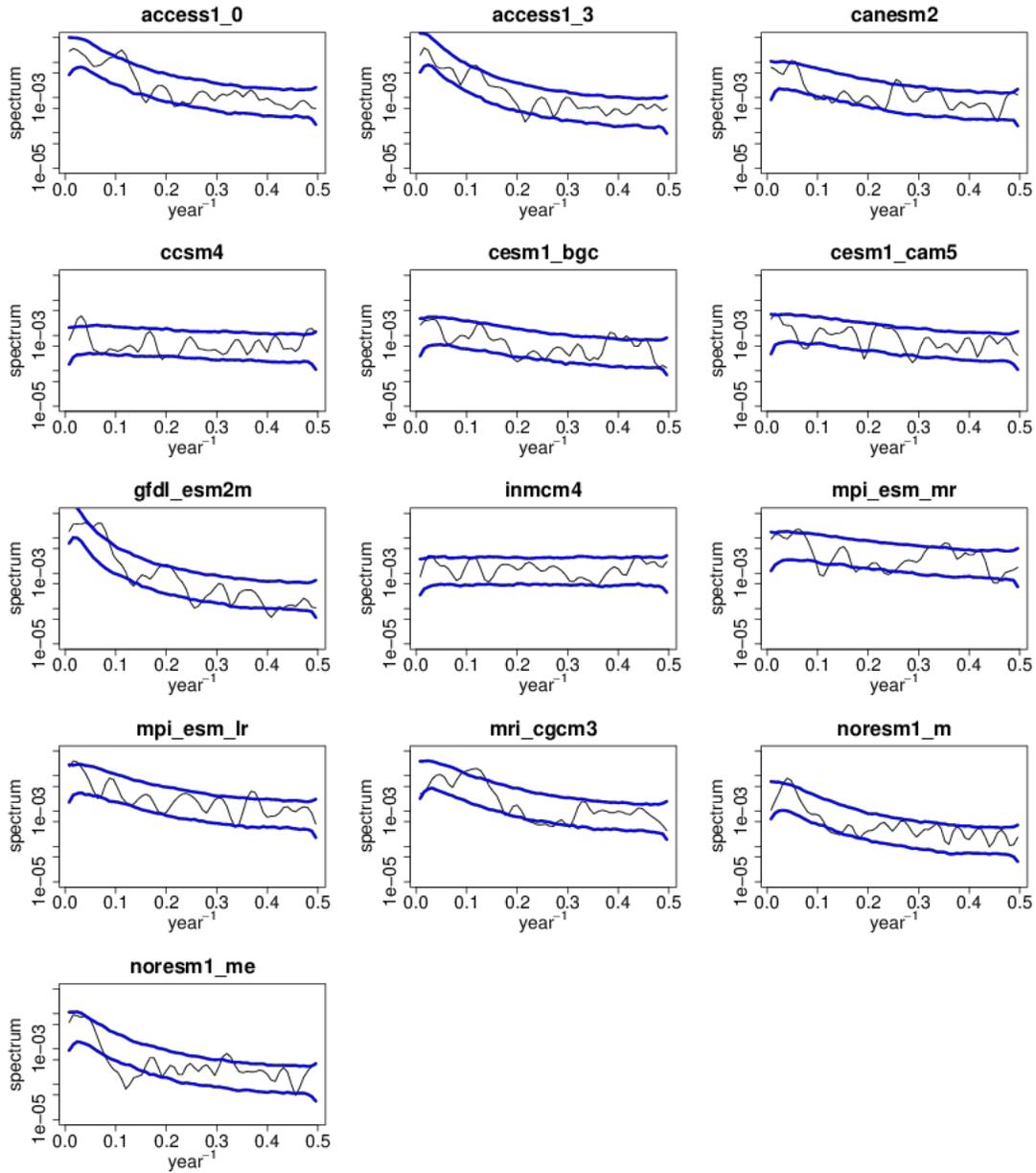

**Fig R.** Spectra plots for AMOC experiment model fluctuations for years 1880-2004. Blue lines: 90% confidence intervals for spectra of an AR1 process that was fit to modelled fluctuations, using 1000 realizations. *Y*-axis is logarithmic.

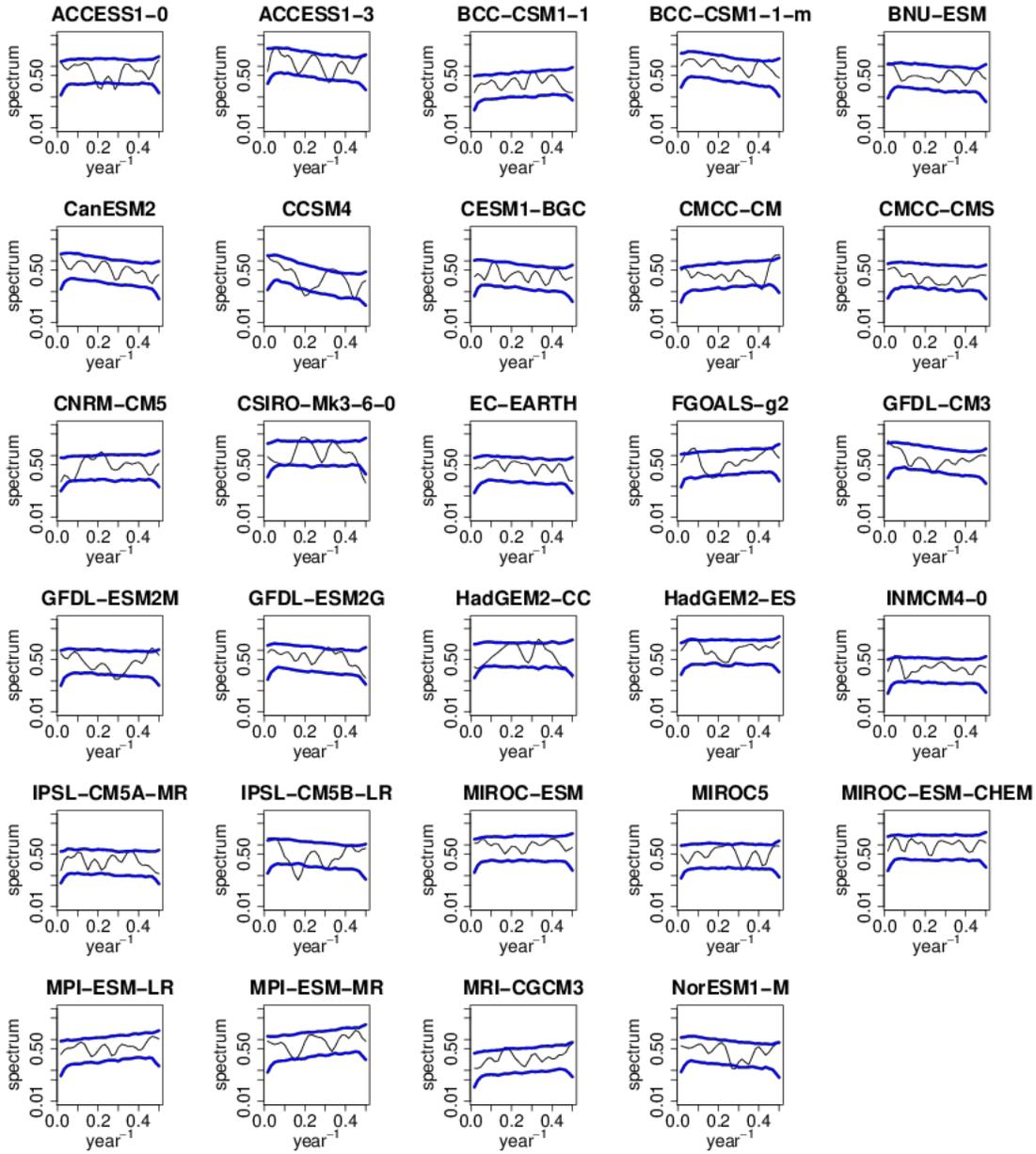

**Fig S.** Same as Fig Q but for Korea_temp_long experiment model fluctuations for years 1950-2005. *Y*-axis is logarithmic.

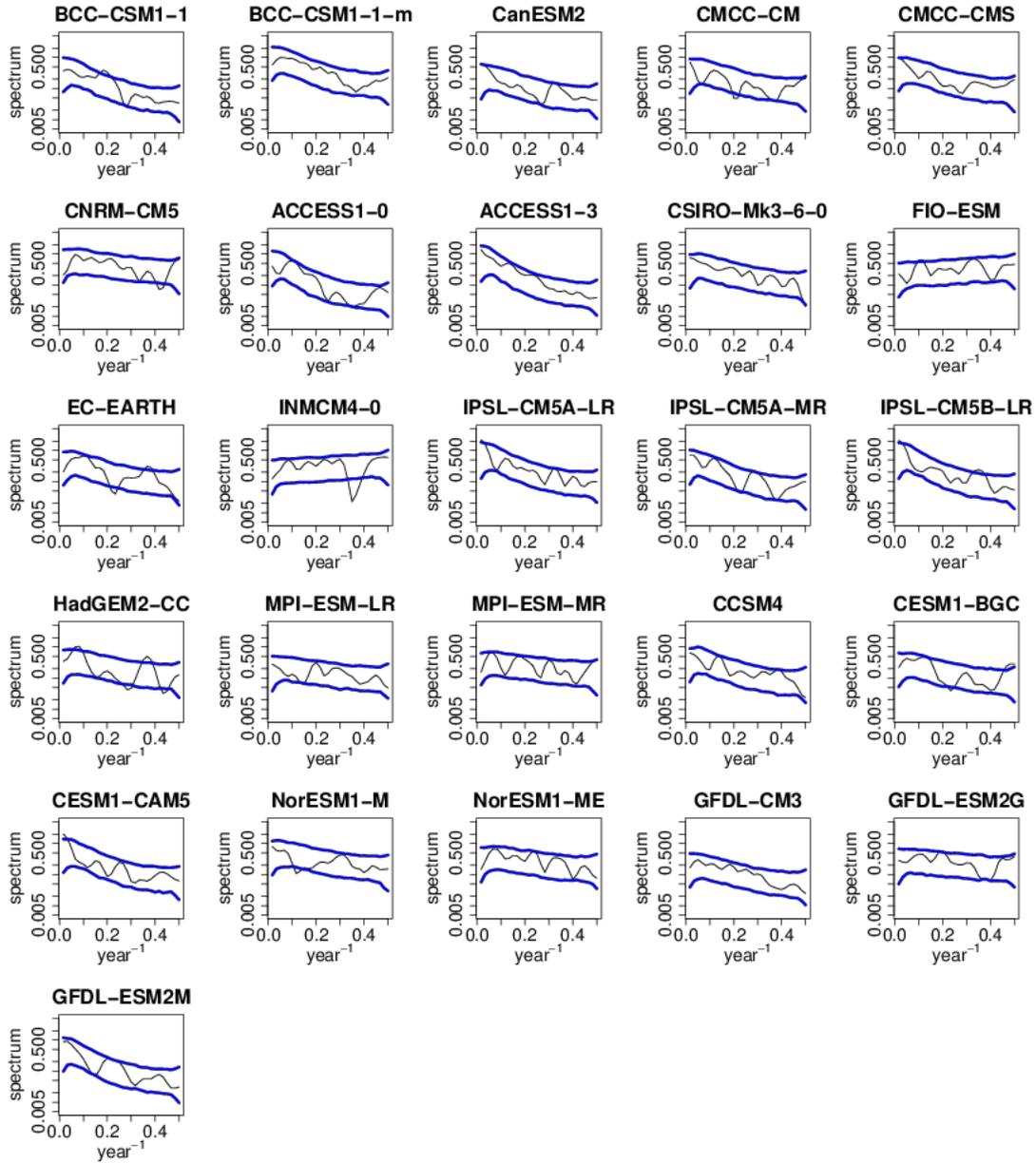

**Fig T.** Same as Fig Q but for Winter_SST_ experiment model fluctuations for years 1941-2000. *Y*-axis is logarithmic.

| Model Number | Name | Modeling Centre |
|---|---|---|
| 1 | ACCESS1-0 | CSIRO and BOM, Australia |
| 2 | ACCESS1-3 | CSIRO and BOM, Australia |
| 3 | bcc-csm1-1 | Beijing Climate Center & China Meteorological Administration, China |
| 4 | bcc-csm1.1-m | Beijing Climate Center & China Meteorological Administration, China |
| 5 | BNU-ESM | Beijing Normal University, China |
| 6 | CanESM2 | Canadian Centre for Climate Modeling and Analysis, Canada |
| 7 | CCSM4 | National Center for Atmospheric Research, USA |
| 8 | CESM1-BGC | National Science Foundation, Department of Energy, National Center for Atmospheric Research, USA |
| 9 | CMCC-CM | Euro-Mediterranean Centre on Climate Change, Italy |
| 10 | CMCC-CMS | Euro-Mediterranean Centre on Climate Change, Italy |
| 11 | CNRM-CM5 | National Centre for Meteorological Research & European Centre for Research and Advanced Training in Scientific Computation, France |
| 12 | CSIRO-Mk3-6-0 | Queensland Centre for Climate Change Excellence & CSIRO, Australia |
| 13 | EC-EARTH | EC-Earth consortium, Europe |
| 14 | FGOALS-g2 | Institute of Atmospheric Physics, Chinese Academy of Sciences, China |
| 15 | GFDL-CM3 | Geophysical Fluid Dynamics Laboratory, USA |
| 16 | GFDL-ESM2M | Geophysical Fluid Dynamics Laboratory, USA |
| 17 | GFDL-ESM2G | Geophysical Fluid Dynamics Laboratory, USA |
| 18 | HadGEM2-CC | Met Office Hadley Centre, UK |
| 19 | HadGEM2-ES | Met Office Hadley Centre, UK |
| 20 | inmcm4-0 | Institute of Numerical Mathematics, Russia |
| 21 | IPSL-CM5A-MR | Institute Pierre Simon Laplace, France |
| 22 | IPSL-CM5B-LR | Institute Pierre Simon Laplace, France |
| 23 | MIROC-ESM | University of Tokyo, National Institute for Environmental Studies & Japan Agency for Marine-Earth Science and Technology, Japan |
| 24 | MIROC5 | University of Tokyo, National Institute for Environmental Studies & Japan Agency for Marine-Earth Science and Technology, Japan |
| 25 | MIROC-ESM-CHEM | University of Tokyo, National Institute for Environmental Studies & Japan Agency for Marine-Earth Science and Technology, Japan |
| 26 | MPI-ESM-LR | Max Planck Institute for Meteorology (MPI-M), Germany |
| 27 | MPI-ESM-MR | Max Planck Institute for Meteorology (MPI-M), Germany |
| 28 | MRI-CGCM3 | Meteorological Research Institute, Japan |

| 29 | NorESM1-M | Norwegian Climate Centre, Norway |

**Table A:** Basic information about GCMs used for the Korea_temp and Korea_temp_long experiments.

| Model Number | Name | Modeling Centre |
|---|---|---|
| 1 | bcc-csm1-1 | Beijing Climate Center & China Meteorological Administration, China |
| 2 | bcc-csm1.1-m | Beijing Climate Center & China Meteorological Administration, China |
| 3 | CanESM2 | Canadian Centre for Climate Modeling and Analysis, Canada |
| 4 | CMCC-CM | Euro-Mediterranean Centre on Climate Change, Italy |
| 5 | CMCC-CMS | Euro-Mediterranean Centre on Climate Change, Italy |
| 6 | CNRM-CM5 | National Centre for Meteorological Research & European Centre for Research and Advanced Training in Scientific Computation, France |
| 7 | ACCESS1-0 | CSIRO and BOM, Australia |
| 8 | ACCESS1-3 | CSIRO and BOM, Australia |
| 9 | CSIRO-Mk3-6-0 | Queensland Centre for Climate Change Excellence & CSIRO, Australia |
| 10 | FIO-ESM | The First Institute of Oceanography, SOA, China |
| 11 | EC-EARTH | EC-Earth consortium, Europe |
| 12 | inmcm4-0 | Institute of Numerical Mathematics, Russia |
| 13 | IPSL-CM5A-LR | Institute Pierre Simon Laplace, France |
| 14 | IPSL-CM5A-MR | Institute Pierre Simon Laplace, France |
| 15 | IPSL-CM5B-LR | Institute Pierre Simon Laplace, France |
| 16 | HadGEM2-CC | Met Office Hadley Centre, UK |
| 17 | MPI-ESM-LR | Max Planck Institute for Meteorology (MPI-M), Germany |
| 18 | MPI-ESM-MR | Max Planck Institute for Meteorology (MPI-M), Germany |
| 19 | CCSM4 | National Center for Atmospheric Research, USA |
| 20 | CESM1-BGC | National Science Foundation, Department of Energy, National Center for Atmospheric Research, USA |
| 21 | CESM1-CAM5 | National Science Foundation, Department of Energy, National Center for Atmospheric Research, USA |
| 22 | NorESM1-M | Norwegian Climate Centre, Norway |
| 23 | NorESM1-ME | Norwegian Climate Centre, Norway |
| 24 | GFDL-CM3 | Geophysical Fluid Dynamics Laboratory, USA |
| 25 | GFDL-ESM2G | Geophysical Fluid Dynamics Laboratory, USA |
| 26 | GFDL-ESM2M | Geophysical Fluid Dynamics Laboratory, USA |

**Table B.** Basic information about GCMs used for the Winter_SST experiment.